\documentclass[aip,rsi,amsmath,amssymb,reprint]{revtex4-1}
\usepackage{graphicx,caption,subcaption}
\usepackage{bm}
\usepackage{hyperref}
\usepackage[capitalise]{cleveref}% noabbrev
\usepackage[normalem]{ulem} %strikethrough 
\usepackage{xcolor} %color
\usepackage{multirow} % for the table

\begin{document}

\title{A machine learning potential for simulating infrared spectra of nanosilicate clusters}

\author{Zeyuan Tang}
\affiliation{Center for Interstellar Catalysis, Department of Physics and Astronomy, Aarhus University, Ny Munkegade 120, Aarhus C 8000, Denmark}

\author{Stefan T. Bromley}
\affiliation{Departament de Ciència de Materials i Química Física \& Institut de Química Teòrica i Computatcional (IQTCUB), Universitat de Barcelona, c/Martí i Franquès 1-11, 08028 Barcelona, Spain}
\affiliation{Institució Catalana de Recerca i Estudis Avançats (ICREA), Passeig Lluis Companys 23, 08010 Barcelona, Spain}

\author{Bjørk Hammer}
\email{hammer@phys.au.dk}
\affiliation{Center for Interstellar Catalysis, Department of Physics and Astronomy, Aarhus University, Ny Munkegade 120, Aarhus C 8000, Denmark}

\date{\today}

\begin{abstract}
The use of machine learning (ML) in chemical physics has enabled the construction of interatomic potentials having the accuracy of ab initio methods and a computational cost comparable to that of classical force fields.
Training an ML model requires an efficient method for the generation of training data.
Here we apply an accurate and efficient protocol to collect training data for constructing a neural network based ML interatomic potential for nanosilicate clusters.
Initial training data are taken from normal modes and farthest point sampling.
Later on, the set of training data is extended via an active learning strategy in which new data are identified by the disagreement between an ensemble of ML models.
The whole process is further accelerated by parallel sampling over structures.
We use the ML model to run molecular dynamics (MD) simulations of nanosilicate clusters with various sizes, from which infrared spectra with anharmonicity included can be extracted.
Such spectroscopic data are needed for understanding the properties of silicate dust grains in the interstellar medium (ISM) and in circumstellar environments.

\end{abstract}

\maketitle

\section{Introduction}
Silicates are the main constituent of dust grains in the interstellar medium (ISM) \cite{henning2010} and in circumstellar environments \cite{tielens1997}.
They provide surfaces for chemical reactions and nucleation sites for the formation of ice mantles \cite{potapov2021}. % and other applications
Silicate dust with different compositions, sizes and shapes exhibit varied properties \cite{min2007} that can be probed by infrared (IR) observations \cite{kemper2004,spoon2022}.
For example, a broad band at 9.7 $\mu$m is normally assigned to Si-O stretching modes and a band around 18 $\mu$m can be attributed to O-Si-O bending modes.
Using an astrophysical model to describe IR emission from carbonaceous and silicate dust grains \cite{draine2001,li2001a},
it has been estimated that around 10\% of interstellar Si could be locked up in ultrasmall silicate grains with radii $<$ 1.5 nm \cite{li2001}.
Therefore, having efficient and accurate methods to model the structures and vibrational properties of nanosilicate clusters is beneficial for understanding IR features and other properties of these, presumably very abundant, species.

Most previously reported spectroscopic data of silicate clusters are from classical force field modelings \cite{zamirri2019} and quantum chemical calculations \cite{escatllar2019,zamirri2019,guiu2021}.
Although producing silicate clusters in experiments is difficult \cite{sabri2013}, recent cluster beam studies have produced small silicate clusters and verified their structures via their theoretically calculated IR spectra \cite{marinosoguiu2022}.
Many theoretical studies of silicates are based on density functional theory (DFT) calculations, whereby the vibrational spectra assume the harmonic approximation.
%However, when silicate clusters exhibit conformational fluxionality at finite temperature, an anharmonic treatment becomes necessary \cite{guiu2021}.
However, at finite temperature, an anharmonic treatment becomes necessary due to a number of possible associated effects (e.g. thermal peak broadening, frequency redshifting, combination bands, conformational fluxionality) \cite{guiu2021}.  
In this case, ab initio molecular dynamics (AIMD) is a more appropriate method to compute IR spectra with anharmonic and temperature effects included \cite{guiu2021}.
Hybrid functionals are often used in IR calculations of nanosilicate clusters in order to provide sufficient accuracy \cite{guiu2021}.
However, the high computational costs of AIMD simulations with hybrid functionals hinder their application for large cluster sizes.
Computationally inexpensive interatomic potentials are needed for MD simulations with large system sizes.
A number of such potentials have been developed and parameterized classical force fields thus exist, that do describe bulk silicate \cite{price1987,walker2003} and silicate clusters \cite{escatllar2019} quite well.
They are often used to roughly predict structure and energetic properties.
% TODO: interatomic potential vs force field, choose one terminology
For example, a force field optimized for silicate clusters has been used for global optimization of silicate clusters with different stoichiometries and sizes \cite{escatllar2019}.
However, classical force fields are sometimes limited by their accuracy.
The low-energy isomers predicted by force fields are thus typically refined by a more accurate method like DFT to ensure an accurate energetic ordering is obtained.
%Classical force fields are fast and used for silicates. % Stefan and other group's silicate force field.
%Few of them are optimized for dipole moments which are gradients to get infrared intensity.

Machine learning (ML) techniques such as artificial neural network \cite{behler2007,ANI-1,SchNet,PhysNet,GM-NN,iGM-NN} and gaussian process regression \cite{bartok2010,koistinen2017,denzel2018a} are powerful tools to fit interatomic potentials \cite{behler2021,deringer2021a} while balancing accuracy and computational costs.
They have been successfully applied to global optimization \cite{ouyang2015,chiriki2017,GOFEE,paleico2020,paleico2020a,timmermann2020,kaappa2021,bisbo2022,christiansen2022,ronne2022}, molecular dynamics \cite{li2015d,caro2018,caro2020,lim2020,noe2020,deringer2021,westermayr2021,unke2021,musaelian2023} and vibrational spectroscopy \cite{gastegger2017,lam2020,FieldSchNet,kaser2021,beckmann2022}.
In this work, we will develop neural network based potentials for silicate clusters.
The desired applications are MD simulations of silicate clusters for variable compositions, sizes and temperatures, from which IR spectra can be extracted.
The paper is organized as follows.
We will first summarize the essential elements to develop an accurate ML potential for three target properties (total energy, atomic forces and total dipole moment) of silicate clusters.
This includes our choice of neural network architectures and the generation of training data.
We move on to a discussion of how the ML potential is trained and validated on the three target properties.
Its spectroscopic predictions are then examined with respect to both harmonic and MD-based spectra.
Afterwards, the transferability of the ML potential is tested on the silicate clusters that are not included in the training data.
Finally, the ML potential is used in MD simulations of silicate clusters and the astrophysical implication of the obtained IR spectra is discussed.

\section{Methods} \label{methods}
% figures
% 1) scheme of normal mode sampling and farthest point sampling
% 2) scheme of adaptive sampling, different structures, temperatures and MD-time curves with uncertainties
In order to compute IR spectra via an MD-based approach, it is needed to have an accurate description of total energy, atomic forces and total dipole moment.
Total energy and atomic forces are trained in one model since atomic forces are the negative gradient of the total energy and the inclusion of forces in training often improves model accuracy for both energy and forces \cite{le2009,bartok2010,artrith2012,christensen2020a}.
The total dipole moment is considered as a property which is independent from the energy and the forces.
As such, it is therefore trained in a separate model following Gastegger et al \cite{gastegger2017}.
Our choice of neural network models is described in details in \cref{NN-model}.
One of the challenges in training such a model is to make sure the trained model is accurate and robust during a long time MD simulations (needed for spectroscopic accuracy).
% TODO (Stefan's comment): 800K for circumstellar environments, we only said interstellar silicate which is much cooler.
The trained model should also be applicable for MD simulations at high temperature (e.g. 800K) which are needed for mimicking some circumstellar conditions \cite{guiu2021,heese2017}.
To address this issue, we will use an active learning strategy \cite{kolsbjerg2018,vanderoord2022} to train the ML model.
The model gets iteratively improved.
During each iteration, the model is used in an MD sampling for exploring the configuration space and selecting training data \cite{behler2015,gastegger2017}.
Both low and high temperatures are chosen in the MD sampling so that different regions of the configuration space are covered.
More details can be found in \cref{training-data}.
Another challenge is the selection of training data which is expected to be representative and unique.
This could also be addressed by the active learning.
At each iteration, only uncertain data that the current model cannot predict well will be added to the training data \cite{kolsbjerg2018,gastegger2017}.
More discussions will be given in \cref{training-data}.
Finally, MD details will be described in \cref{MD} and DFT settings will be summarized in \cref{DFT}.

\subsection{Neural network models} \label{NN-model}
The Gaussian moment neural network (GM-NN) \cite{GM-NN,iGM-NN} developed by Kästner's group was used to construct potentials for total energies and atomic forces.
The total energy of a given system $\hat{\text{E}}$ is calculated as a sum of atomic energies $\epsilon_i$.
\begin{equation}
	\hat{\text{E}} = \sum_{i=1}^{\text{N}_\text{atoms}} \epsilon_i
\end{equation}
Those atomic energies are the outputs from GM-NN models which take the local atomic environments (learnable Gaussian moments in this model) as inputs.
The prediction of atomic force $\hat{\textbf{F}}_i$ on an atom $i$ is achieved by taking the partial derivative of the total energy with respect to the atomic position $\textbf{r}_i$ of atom $i$.
\begin{equation}
	\hat{\textbf{F}}_i = - \dfrac{\partial \hat{\text{E}}}{\partial \textbf{r}_i}
\end{equation}
During the training, a combined loss function is used.
\begin{equation}
	\mathcal{L}_\text{E,F} = \sum_{n=1}^{N_\text{structures}} \biggl[ \lambda_\text{E} || \text{E}^\text{ref} - \hat{\text{E}} ||^2 + \frac{\lambda_\text{F}}{3\text{N}_\text{atoms}} \sum_{i=1}^{\text{N}_\text{atoms}} || \textbf{F}^\text{ref}_i - \hat{\textbf{F}}_i ||^2 \biggr]
\end{equation}
where the trade-off between energy and force is set to $\lambda_\text{E} = 1$ and $\lambda_\text{F} =$ 10 Å$^2$.
The GM-NN code \cite{GM-NN-code} was used to train total energies and atomic forces.

The dipole moment was trained separately since the GM-NN code \cite{GM-NN-code} cannot provide dipole moment predictions yet. 
We used the SchNet neural network \cite{SchNet} implemented in SchNetPack v1.0.0 \cite{SchNetPack} to model the dipole moment.
The total electric dipole moment is computed as a sum of atomic charges weighted by their positions with respect to the center of mass.
\begin{equation}
	\bm{\hat{\mu}} = \sum_{i=1}^{\text{N}_\text{atoms}} \hat{q}_i (\textbf{r}_i - \textbf{r}_0)
\end{equation}
where $\hat{q}_i$ is the charge of atom $i$ predicted by a SchNet model and $\textbf{r}_0$ is the center of mass for a given structure.
Note that atomic charges are not fixed for each individual atom, but depend on their atomic environments.
Consequently, atomic charges are fluctuating in MD simulations.
The loss function for dipole moment is defined as 
\begin{equation}
	\mathcal{L}_\mu = \sum_{n=1}^{N_\text{structures}} \biggl[ \frac{1}{3} || \bm{\mu}^\text{ref} - \hat{\bm{\mu}} ||^2 \biggr]
\end{equation}
\subsection{Training data} \label{training-data}
The silicate clusters investigated in this work are taken from global minimum energy structures of pyroxene (MgSiO$_3$)$_\text{N}$ and olivine (Mg$_2$SiO$_4$)$_\text{N}$ with a range of cluster sizes (N=1-10) \cite{escatllar2019}.
The name of those silicate clusters is abbreviated as P1, P2, ... and P10 for pyroxene (MgSiO$_3$)$_\text{N}$ (N=1-10).
Similarly, O1, O2, ... and O10 stand for olivine (Mg$_2$SiO$_4$)$_\text{N}$ (N=1-10).
Their structures are shown in Fig. S1.
During the construction of ML interatomic potentials, the generation of training data is an important factor next to choosing appropriate ML models.
The ideal training data should be representative and not redundant in the configuration space.
We have used an active learning strategy to generate our training data \cite{gastegger2017,kolsbjerg2018,gastegger2020}.
The active learning has been widely used to generate training data while retaining small amounts of training data \cite{smith2018,bernstein2019,schran2020,vandermause2020,miksch2021a,xie2021,young2021}.

\begin{figure}
	\begin{center}
	\includegraphics[width=\columnwidth]{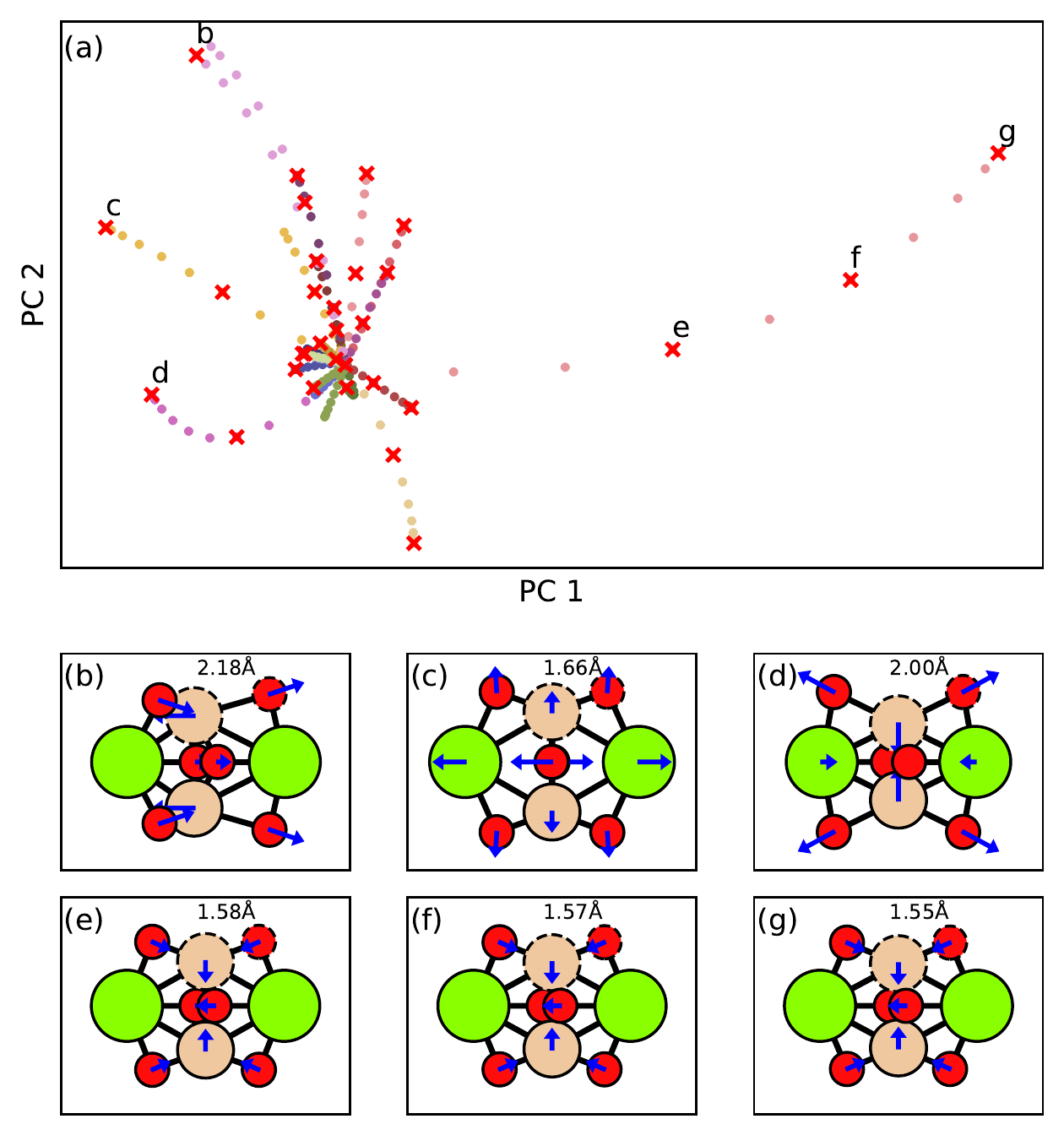}
	\caption{
		Farthest point sampling on displaced structures along normal modes.
		(a) Principal component analysis (PCA) on the global features of various P2 structures.
		The filled circles in the same color are the structures along one specific mode and different color indicates different modes.
		The red crosses are the selected points from farthest point sampling.
		(b) - (g) Snapshots of the selected structures.
		The normal vectors are plotted in blue arrows.
		One Si-O bond length is labeled to quantify the difference between those six structures.
	}
    \label{fig:initial-sampling}
    \end{center}
\end{figure}

The first step is to make an initial set of training data.
Initial training data are generated by normal mode sampling \cite{ANI-1,braams2009,qu2018} with a modification.
Each normal mode represents a unique way to displace a structure from its local minimum as seen in \cref{fig:initial-sampling}.
All displaced structures are further filtered by farthest point sampling (FPS) \cite{bartok2017,imbalzano2018} in order to keep the size of training data small.
%For each silicate cluster, $3\times \text{N}_\text{atoms} + 1$ structures are selected in which one is the local minimum and $3\times \text{N}_\text{atoms}$ are filtered structures along normal modes.
For each silicate cluster, one local minimum structure is selected along with $3\times \text{N}_\text{atoms}$ structures from our modified normal mode sampling.
In the end, 2000 structures collected from all silicate systems are used to train an initial ML model.
The whole training data set is split into a training subset and a validation subset with a ratio of 9:1 during initial and subsequent trainings.

\begin{figure*}
	\begin{center}
	\includegraphics[width=\textwidth]{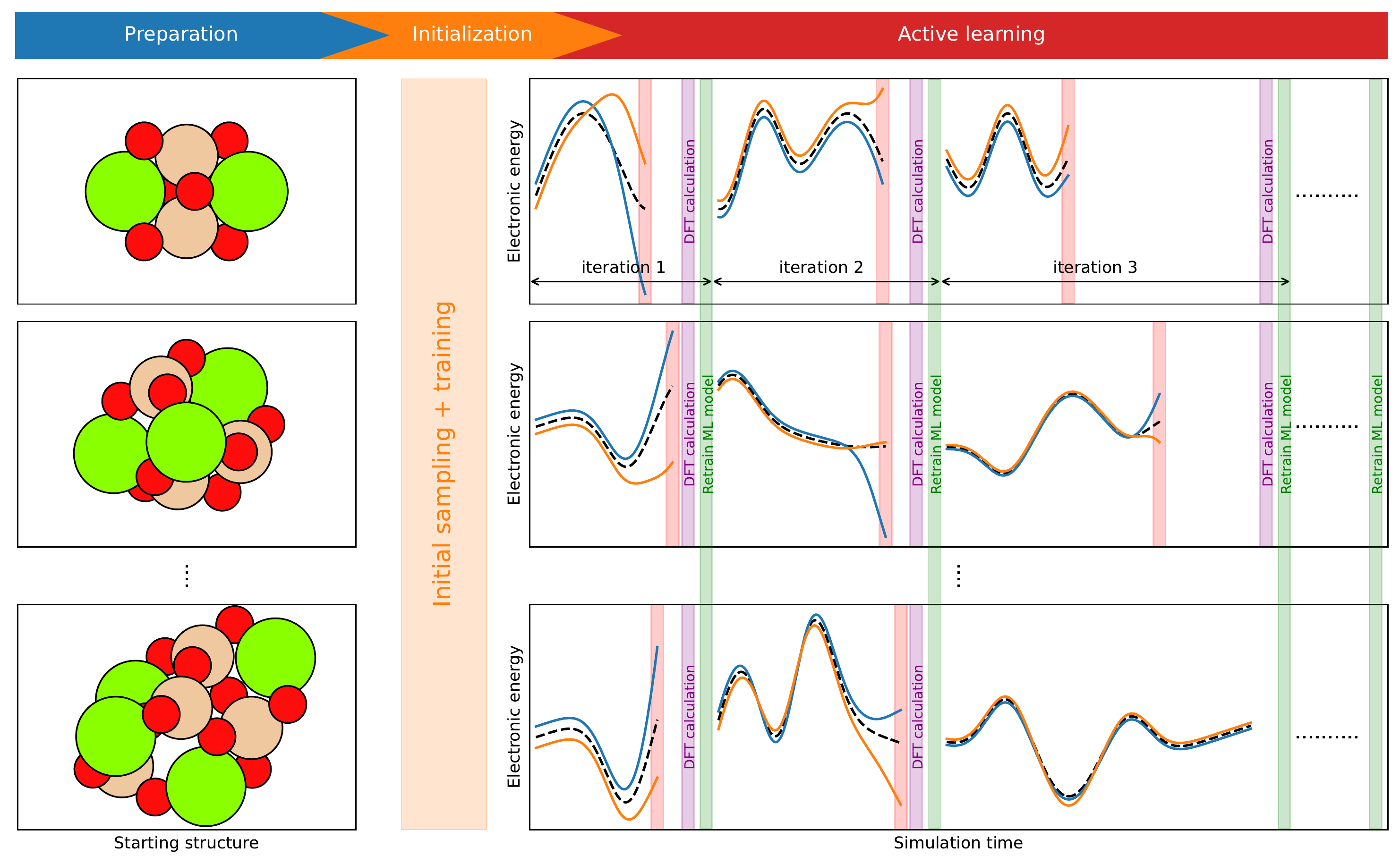}
	\caption{
		Scheme of the active learning in a parallel manner.
		In each sampling panel, blue and orange lines are energy predictions from two independent ML models trained on the same data.
		Dashed black lines are the target DFT potential.	
		Each vertical red bar indicates the appearance of an uncertain data.
		%The scheme is just for illustration purpose and the data shown here are not from real simulations.
	}
    \label{fig:adpative-scheme}
    \end{center}
\end{figure*}

Later on, the ML model is iteratively improved by adding training data from the active learning following the scheme in \cref{fig:adpative-scheme}.
The sampling is driven by MD using the ML model and is paused upon the appearance of uncertain data that the current ML model cannot describe well.
A straightforward approach is to run a DFT calculation at each MD step and compare with the model prediction.
Uncertain data can be easily located with this method which is, however, computationally expensive. 
We use the query by committee method \cite{smith2018,schran2020} which is a common approach to get an uncertain estimate without heavy DFT calculations.
It uses an ensemble of ML models and finds uncertain data where different ML models disagree with each other regarding the energy prediction \cite{schran2019,schran2020}.
The uncertain data is recomputed at the DFT level and added to the training data.
A new ML model is trained after the collection of uncertain data.
The active learning then moves to next iteration in which MD sampling is performed again.
The MD trajectory from previous iteration is used as the starting geometry for the following iteration.
For each new iteration, the distribution of atomic velocities is also taken from the previous iteration, and the overall translation and rotation are removed.
%Its velocity distributions are loaded as well, but overall linear and angular momenta are removed.
The target simulation temperature remains unchanged from one iteration to another.

In order to further speed up the sampling, multiple systems are sampled in parallel at different temperatures and stoichiometries \cite{gastegger2020}.
When samplings are finished, DFT calculations on uncertain data can be parallelized as well.
Therefore, multiple structures (from one to twenty) are added to the training data in one iteration.
Only one job of ML training in one iteration is conducted using the updated training data.

In summary, our protocol for collecting training data was composed of iterations with these steps as shown in \cref{fig:adpative-scheme}:
\begin{enumerate}
	\item Parallelize over structures
	\begin{enumerate}
		\item MD propagation until the appearance of an uncertain data or reaching the maximum simulation time
		\item Store uncertain data if appeared
	\end{enumerate}	 
	\item Wait until all MD runs have ended
	\item Perform DFT calculations on uncertain data in a massively parallel manner
	\item Train the ML model
\end{enumerate}
The collection of training data will stop when the accuracy of the ML model cannot be improved by adding more training data.
In practice, we assess the difference between DFT and ML in terms of harmonic frequencies of all investigated silicate clusters.
We stop collecting training data when this difference cannot be decreased within a number of iterations.
Finally, 1458 instances of uncertain data are collected and the final training data consists of 3458 structures.

\subsection{MD simulations} \label{MD}
For all MD simulations, a time step of 0.5 fs was used.
During the active learning, the temperature was controlled by a Berendsen thermostat \cite{berendsen1984} with a time constant of 100 fs implemented in ASE 3.21.1 \cite{ASE}.
The maximum simulation time was set to 5 ps during the active learning.
During the production runs for the IR spectra, the Nosé-Hoover chain thermostat \cite{martyna1992} in SchNetPack 1.0.0 \cite{SchNetPack} was used to keep a constant temperature.
The time constant was 100 fs and a chain length of 3 was used.
The total simulation time was set to 40 ps.
The first 5 ps was discarded when extracting IR spectra from the autocorrelation function of the dipole time derivative \cite{thomas2013} according to \cref{eq:autocorr}.
\begin{equation}\label{eq:autocorr}
	I_{IR} \propto \int _ {-\infty} ^ {+\infty} \langle	\dot{\mu}(\tau) \dot{\mu}(\tau+t) \rangle_{\tau} e^{-i \omega t} dt
\end{equation}
All autocorrelation functions were computed using the Wiener-Khinchin theorem \cite{wiener1930}.
In order to obtain IR spectra with high quality, a Hann window function \cite{blackman1958} and zero-padding were applied to the autocorrelation functions before the Fourier transform.
A maximum correlation depth of 2048 fs was used.
All processing of IR spectra was done with SchNetPack 1.0.0 \cite{SchNetPack}.

\subsection{DFT simulations} \label{DFT}
All DFT calculations were performed with ORCA 5.0.0 \cite{ORCA5}. 
A hybrid functional PBE0 \cite{PBE0} was used due to its excellent accuracy in modeling structural and spectroscopic properties of silicate clusters \cite{guiu2021}.
A double-zeta basis set def2-SVP \cite{def2} and an auxiliary basis def2/J \cite{def2-fit} were used.
All hybrid DFT calculations were accelerated by the RIJCOSX approximation \cite{RIJCOSX}.
Here, the RI-J approximation is the resolution of identity (RI) approximation for Coulomb integrals (J) \cite{neese2003}.
The "chain of spheres" COSX approximation is used to speed up HF exchange integrals \cite{RIJCOSX}.
For all DFT calculations, the SCF convergence criteria were set to tight.
Default DFT and COSX grids "DEFGRID2" were used, and normally show small errors for energies, geometries and frequencies compared with a larger grid setting.
Examples of ORCA inputs for single point calculations, structure optimization and harmonic calculations are given in the supplementary material.
% ORCA's manual P332: the new default grid settings in ORCA 5.0 (defgrid2) should be sufficient in most cases, certain cases might need the use of defgrid3.
% ORCA's manual P488: In general, the errors from the default grids are rather small and reasonable for most applications.

%\begin{equation}
%	I_{\omega} \propto \int _ {-\infty} ^ {+\infty} \langle	\dot{\mu}(\tau) \dot{\mu}(\tau+t) \rangle_{\tau} e^{-i \omega t} dt
%\end{equation}
\section{Results}
% figures
% 1) sampling strategy, errors during training
% 2) comparing harmonic frequencies of O1-O10 and P1-P10, energy+force for 100ps trajectories
% 3) dipole predictions of final ML model
% 4) comparing AIMD runs of O2, P2 and P5 with ML-MD runs, discuss their band positions, intensities
% 5) more independent runs of O2, P2 and P5
% 6) test harmonic frequencies on other high-energy isomers (do we need AIMD runs for some other isomers?)
%		one example, snapshot with relative energy, harmonic spectra, parity plot for freq and intensity
%		all pyroxene, parity plot for freq and intensity, different colors and alphas
%		all olivine, parity plot for freq and intensity, different colors and alphas
\subsection{Validation of the ML model} \label{validation}
\begin{figure}
	\begin{center}
	\includegraphics[width=\columnwidth]{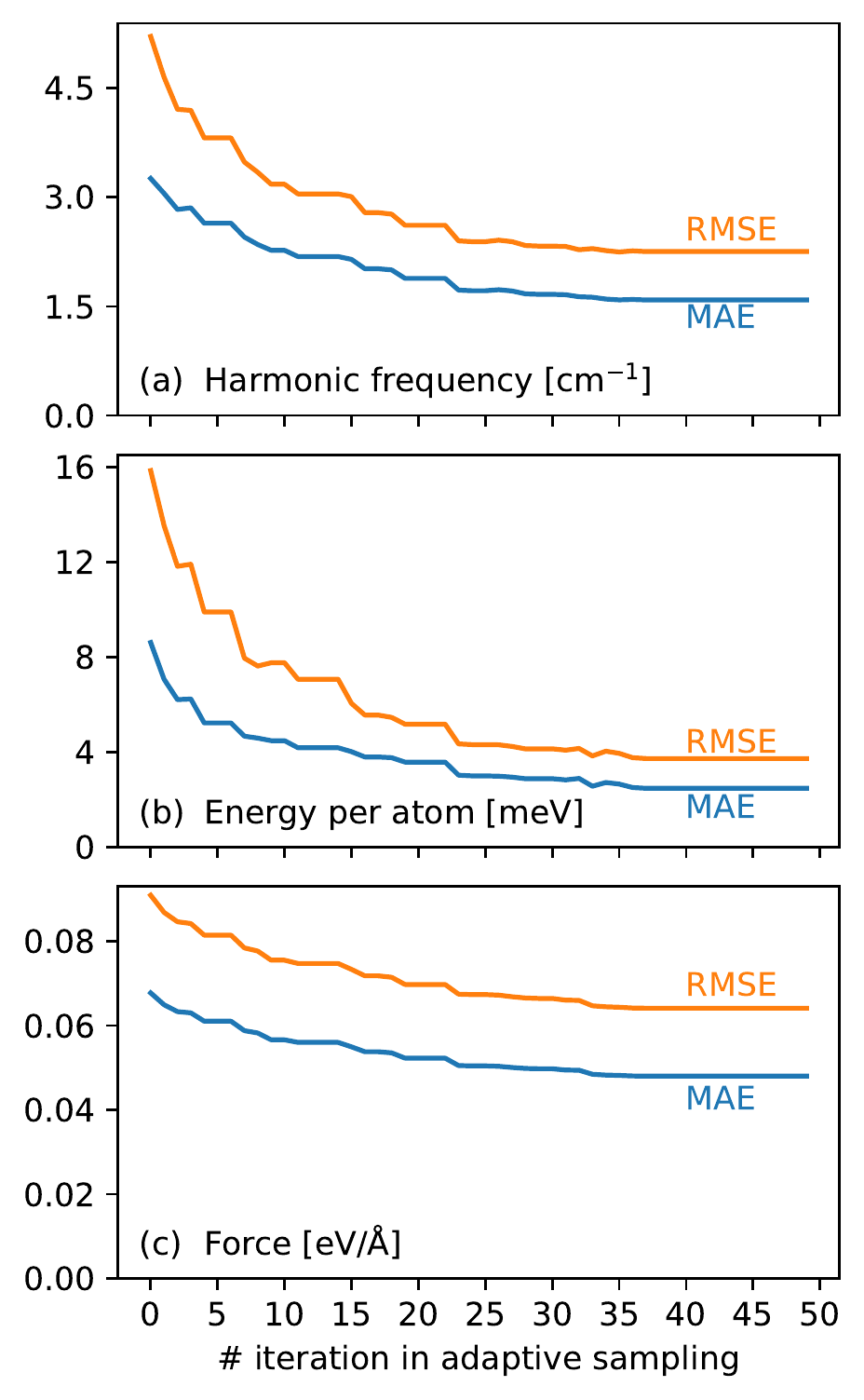}
	\caption{
		Accuracy of ML models at each iteration of active learning, in terms of 
		mean absolute errors (MAEs) and root mean squared errors (RMSEs) of (a) harmonic frequency, (b) energy per atom, and (c) atomic force.
		Harmonic frequency is based on 20 silicate clusters (P1-P10 and O1-O10).
		Energy and forces are based on the test data.
		Note that the test data are not included in the training and are used to assess the accuracy of ML models.
		The generation of test data is described in details in \cref{validation}.
	}
    \label{fig:errors-training}
    \end{center}
\end{figure}

To assess the quality of the trained ML model during the active learning, the model was stored at the end of each iteration and tested against harmonic frequencies of all investigated silicate clusters.
Harmonic frequencies are chosen since the reference frequencies from DFT calculations are already available before the active learning and it is computationally cheap to compute harmonic frequencies with the ML potential.
The mean absolute error (MAE) and root mean squared errors (RMSE) are used to quantify the difference between the results from DFT and ML. 
\cref{fig:errors-training} (a) shows MAE/RMSE of harmonic frequency at each sampling iteration.
Before the active learning, when the training data are only sampled from normal modes, the MAE of harmonic frequencies is 3.2 cm$^{-1}$.
This error decreases with the use of active learning, indicating the improvement of the ML model when more uncertain data are included in the training.
After about 40 iterations, the MAE becomes stagnated and the ML model cannot be further improved by adding more data according to our sampling strategy.
Therefore, the active learning is stopped with a 1.6/2.2 cm$^{-1}$ MAE/RMSE of harmonic frequency.
% harmonic frequencies are evaluated by finite difference, and using DFT-local-minima since ML-opt may failed for some models.
We also need to see if the final ML model can be reliably used in MD simulations.
To assess this, another set of test data other than harmonic frequencies is constructed based on the MD simulations using the final ML model.
The MD simulations are performed for 100 ps for each silicate cluster at three different temperatures (400K, 800K and 1200K).
No obscure configurations (e.g. overlapping atoms and atoms being far from each other) were observed in the trajectory of any above-mentioned MD run.
DFT calculations at the PBE0/def2-SVP level are performed on 100 randomly selected structures from each MD run.
Finally, a test data set consisting of 6000 structures was collected.
To assess the degree of improvement in terms of energy and force predictions, energies and forces are recomputed on the test data using the ML model at each iteration.
Their corresponding MAEs and RMSEs are shown in \cref{fig:errors-training} (b) and \cref{fig:errors-training} (c), and decrease as the iteration proceeds.
%The improvement of energy and force prediction is consistent with that of harmonic frequency when using adaptive sampling.
The final ML model gives an MAE/RMSE of 2.3/3.5 meV for energy per atom and 0.043/0.060 eV/Å for atomic force, based on the test data set.
The accuracy of the final ML model on the training and validation data set are included in \cref{tab:ml-accuracy}.

A comparison between the ML model and classical force fields is given in Fig. S2 in terms of energy and force predictions.
For the energy comparison, we use relative energies with respect to the energy of global minimum structure, divided by the number of atoms.
The force field by Escatllar et al \cite{escatllar2019} is chosen since it is parameterized with respect to relative energies and cluster geometries.
Our ML model shows a one order of magnitude improvement for energy prediction over the Escatllar force field \cite{escatllar2019}.
For the force prediction, we choose the force field by Walker et al \cite{walker2003}, as it is designed to study oxygen diffusion in olivine and more suitable for force prediction than the Escatllar force field \cite{escatllar2019}.
Our ML model shows a two order of magnitude improvement for force prediction over the Walker force field \cite{walker2003}.

\begin{table}
	%\centering
	\caption{Accuracy of the final ML model on training, validation and test data sets} \label{tab:ml-accuracy}
	%\resizebox{\textwidth}{!}{%
	{\tiny
	\begin{tabular}{lllllllll}
	\multirow{2}{*}{Property} & \multirow{2}{*}{Unit} & \multirow{2}{*}{Model} & \multicolumn{2}{l}{Training set} & \multicolumn{2}{l}{Validation set} & \multicolumn{2}{l}{Test set} \\
							  &                       &                        & MAE             & RMSE           & MAE              & RMSE            & MAE           & RMSE         \\
	Energy per atom           & meV                   & GM-NN                  & 1.6             & 3.2            & 1.7              & 3.0             & 2.3           & 3.5          \\
	Atomic forces             & ev/Å                  & GM-NN                  & 0.037           & 0.050          & 0.045            & 0.077           & 0.043         & 0.060        \\
	Dipole moment             & Debye                 & SchNet                 & 0.049           & 0.064          & 0.090            & 0.128           & 0.074         & 0.102       
	\end{tabular}
	}
	%}
\end{table}

\begin{figure}
	\begin{center}
	\includegraphics[width=\columnwidth]{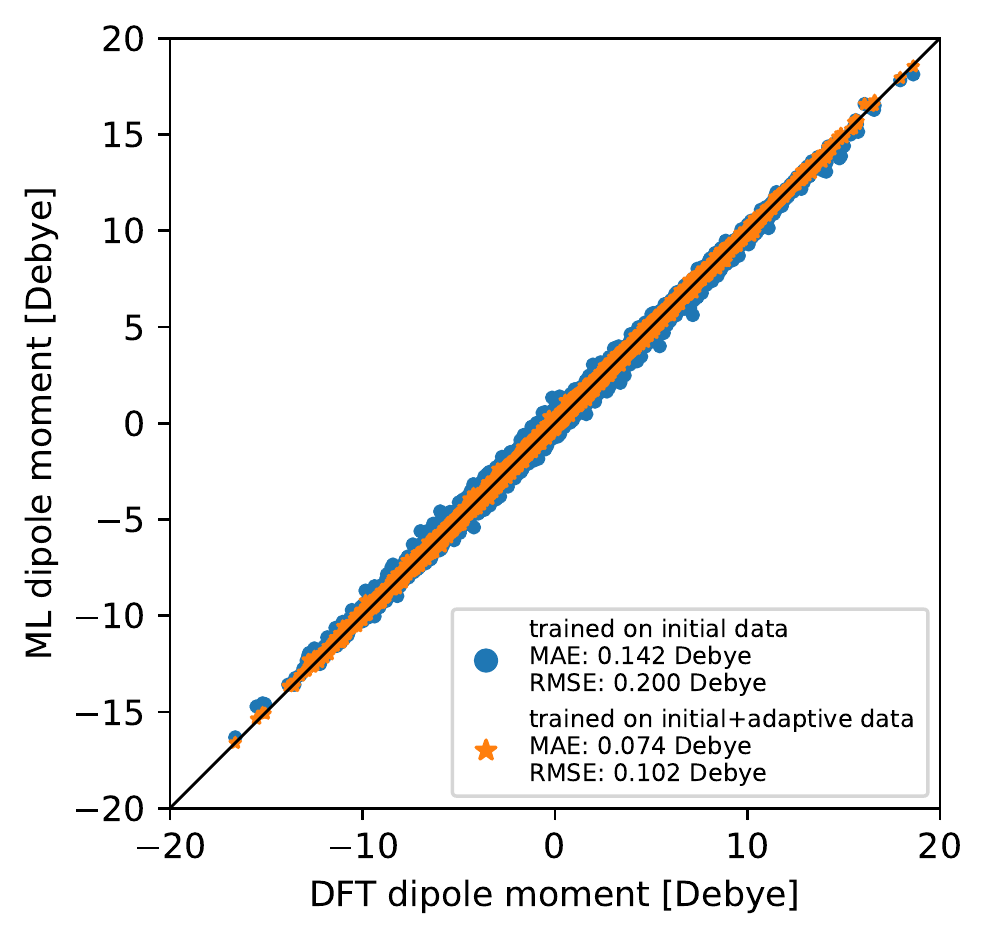}
	\caption{Accuracy of ML models in predicting dipole moment for the test data when different training data are used.
	Note that the test data are not included in the training and are used to assess the accuracy of ML models.
	The generation of test data is described in details in \cref{validation}.
	}
    \label{fig:errors-training-dipole}
    \end{center}
\end{figure}

% We have further tested dipole accuracy on the test data.
% We did not test on dipole moment at each adaptive iteration, but tested on initial and final dataset.
In order to compute infrared intensity, the prediction of dipole moment is needed.
We have trained SchNet models only for dipole moment based on two set of training data, namely the initial dataset and the final dataset after the active learning is finished.
%One data with high dipole component was removed from the training data, since its inclusion lowers the model accuracy in predicting dipole moment.
\cref{fig:errors-training-dipole} compares the reference DFT dipole moment against the predicted dipole moment for the test data when different training data are used.
The ML model shows good accuracy in predicting dipole moment with MAE of 0.142 Debye and RMSE of 0.200 Debye when only structures sampled from normal modes are included in the training.
The ML model from the active learning results in a better accuracy in predicting dipole moment and the MAE/RMSE is only 0.074/0.102 Debye.
Adding training data by the active learning indeed improve the ML model's accuracy regarding dipole moment prediction.

\subsection{Spectroscopic behavior of the ML model}
Our ML models have reproduced the potential energy surface (PES) and dipole surface of DFT to a reasonable degree of accuracy.
The next step is to benchmark the spectroscopic behavior of ML models against DFT results.
IR spectra containing both band positions and infrared intensities at the harmonic level are evaluated first.
Infrared intensities, which are not examined in \cref{validation} yet, are calculated from dipole derivatives along normal modes.
The DFT-calculated band positions and intensities are reproduced almost exactly by ML models for all investigated silicate clusters, see Fig. S3.
MAEs of harmonic frequencies for most clusters are below 2.5 cm$^{-1}$.
P1 and O1 show slightly higher MAEs of 4.2 cm$^{-1}$ and 3.6 cm$^{-1}$.
For harmonic intensities, the averaged MAE for all clusters is 26.8 km/mol.
P2 shows the lowest MAE of 2.8 km/mol and P10 shows the highest MAE of 49.4 km/mol.

\begin{figure*}
	\begin{center}
	\includegraphics[width=\textwidth]{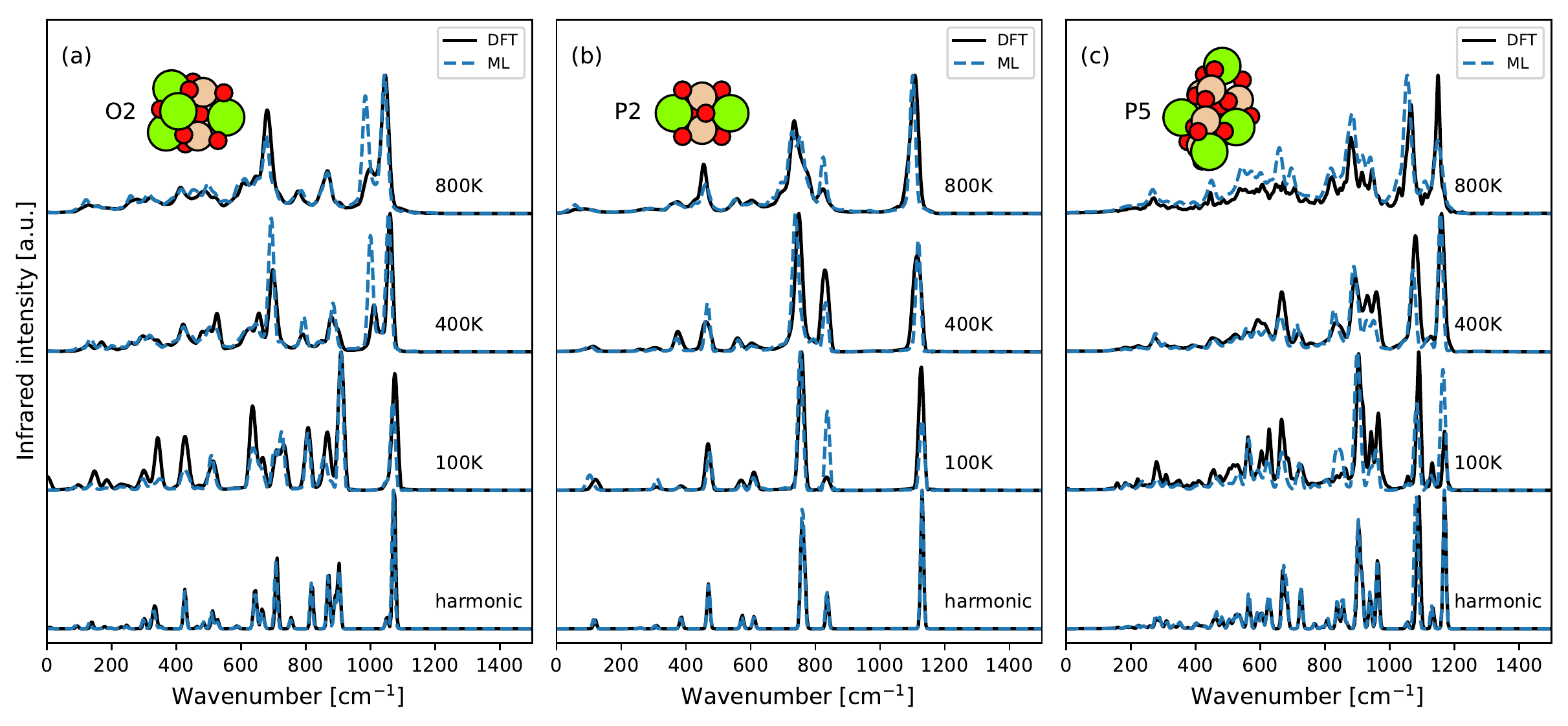}
	\caption{
		Comparing IR spectra from DFT and the ML model.
		Both harmonic and MD-based spectra are given for selected silicate clusters (a) O2, (b) P2, (c) P5 at 100K, 400K and 800K.}
    \label{fig:IR-DFT-ML-MD}
    \end{center}
\end{figure*}

\begin{figure}
	\begin{center}
	\includegraphics[width=\columnwidth]{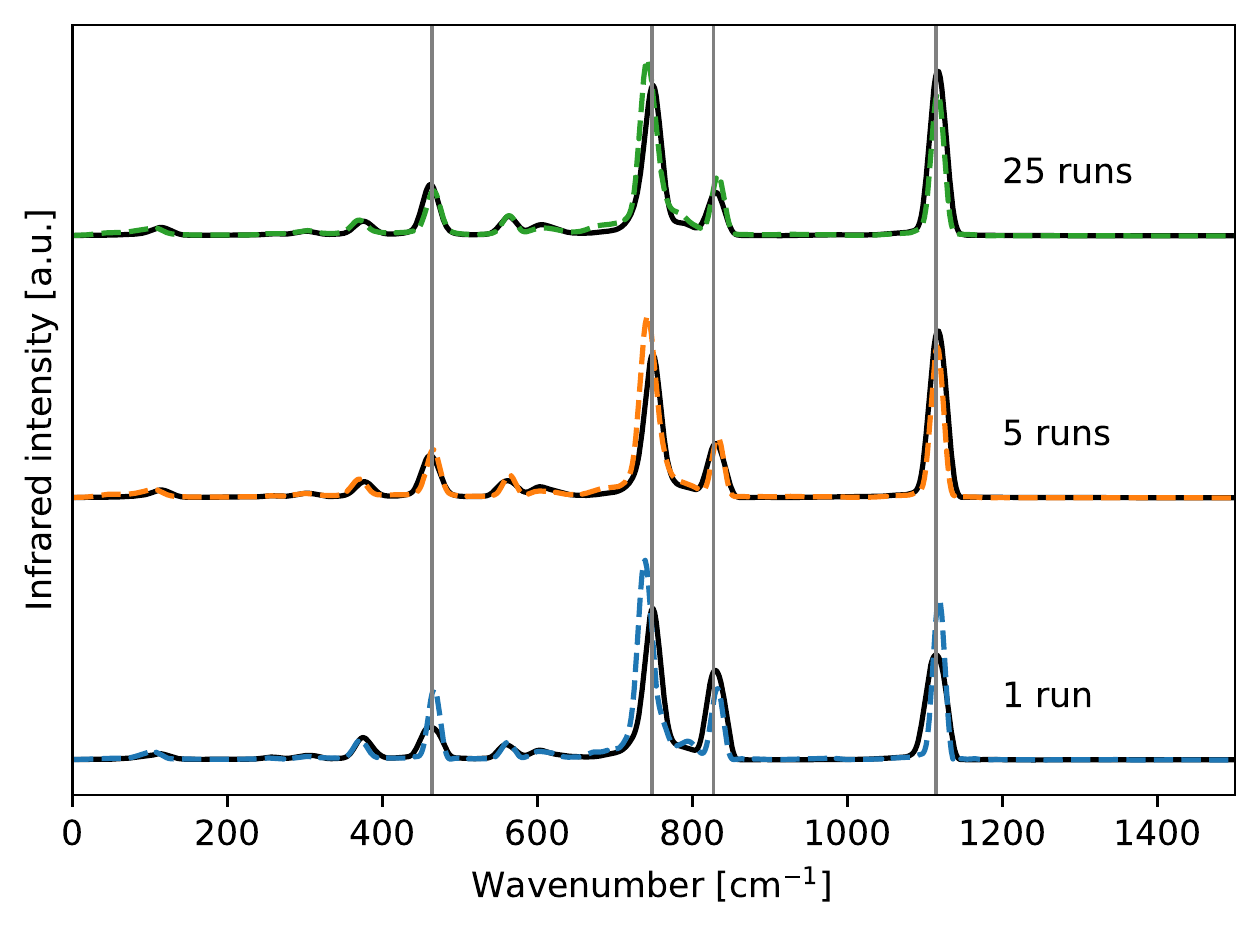}
	\caption{
		MD-based IR spectra of P2 at 400K averaged over independent runs.
		DFT results are shown in solid black lines and ML results are shown in colored dashed lines.
		Number of independent DFT-MD and ML-MD runs is labeled on the right of each spectrum.	
		Bands at 464, 748, 827, and 1114 cm$^{-1}$ are highlighted by vertical gray lines.
	}
    \label{fig:IR-DFT-ML-MD-P5-avg}
    \end{center}
\end{figure}

Selected harmonic spectra are given in the bottom panel of \cref{fig:IR-DFT-ML-MD} along with MD-based spectra at three selected temperatures (100K, 400K and 800K).
For MD-based spectra, band positions are well reproduced by ML models while intensities have larger variations than band positions.
The variations in intensities could arise from differences in velocity initializations in the MD runs or inaccuracy of the ML models.
In order to check the influence of variable velocity initializations, we performed twenty five independent DFT-MD runs for P2 at 400K.
The averaged IR spectra over independent DFT-MD runs are compared with those of ML-MD runs in \cref{fig:IR-DFT-ML-MD-P5-avg}.
When only one MD run is used, intensity variations are observed for peaks with frequencies around 464, 748, 827, and 1114 cm$^{-1}$.
The variation becomes smaller for bands at 464, 827 and 1114 cm$^{-1}$ when the number of independent MD runs increased from one to five, and barely changes when twenty five independent runs are used.
The disagreement between DFT and ML results of band position and intensity at 748 cm$^{-1}$ (Si-O stretching modes between the central O atoms and Si atoms) is not sensitive to the number of independent MD runs.
This disagreement is therefore likely due to model fitting.
Performing independent DFT-MD runs for each silicate size and temperature is computationally expensive.
Therefore only one DFT-MD run and one ML-MD run are compared in \cref{fig:IR-DFT-ML-MD} for each system.
In Fig. S4, one DFT-MD run is compared against independent ML-MD runs.
In general, DFT results are within the range of independent ML runs.
We expect better agreement between DFT and ML when more independent runs are performed.
Since it is computationally expensive to run many DFT-MD simulations, the resulting IR spectra are expected to be biased by the velocity initializations.
However, the use of an ML potential allows for a fast evaluation of many independent MD runs and thus the generation of converged IR spectra.
%We select P5 as the largest system for DFT-MD since it is too expensive to run DFT-MD for all cluster sizes.

\subsection{Transferability of the ML model}
\begin{figure}
	\begin{center}
	\includegraphics[width=\columnwidth]{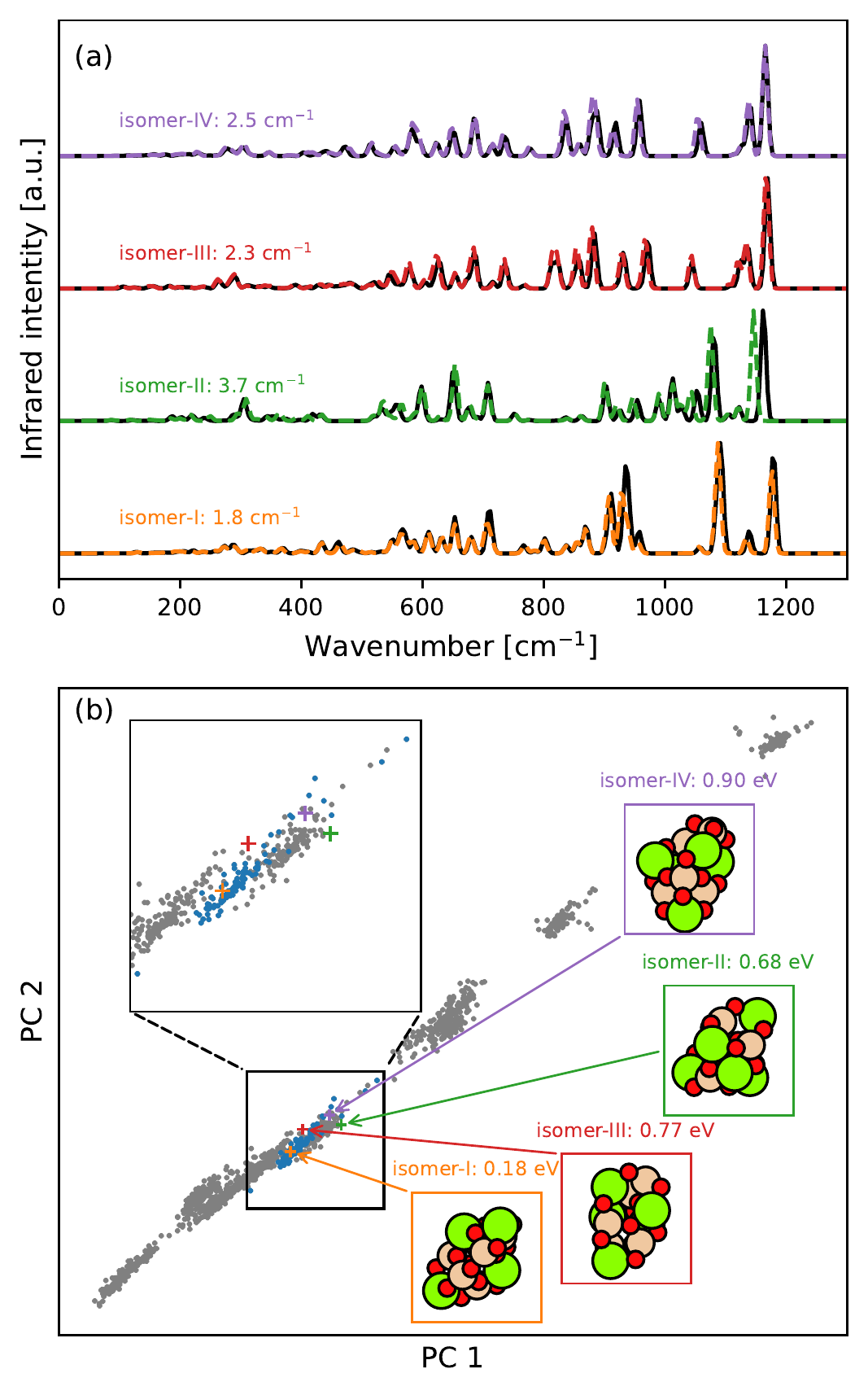}
	\caption{
		Transferability of the ML model on P5 isomers found via the GOFEE method \cite{bisbo2022}.
		(a) Harmonic IR spectra of selected P5 isomers with DFT results in black solid lines and ML results in colored dashed lines.
		MAEs of harmonic frequencies are labeled for each isomer.
		(b) Principal component analysis (PCA) on global Smooth Overlap of Atomic Positions (SOAP) features \cite{SOAP} of all pyroxene clusters.
		Gray circles are training data.
		Blue circles are P5 training data.
		Colored crosses are high-energy P5 isomers with their structures and relative energies to P5 global minimum attached.
	}
    \label{fig:ir-P5-isomers}
    \end{center}
\end{figure}

After the assessment of the ML model, we want to see whether the ML model is accurate and transferable for the structures that are different from the global minimum based training and test data. 
Examples of such structures are high-energy isomers that could be obtained during global optimization searches via the GOFEE (global optimization with first-principles energy expressions) method \cite{bisbo2022}.
At first, P5 high-energy isomers are chosen to test the transferability of the ML model.
Only harmonic spectra are used to verify the transferability of the ML model, since harmonic frequencies are sensitive to the quality of the PES and harmonic calculations at the DFT level are computationally much cheaper than DFT-MD simulations.
In addition, harmonic frequencies are the main reservoir of vibrational transitions for a given nanosilicate cluster.
The synthetic IR spectra of various nanosilicate clusters based on the harmonic approximation are already used to estimate the abundance of nanosilicate clusters in the diffuse ISM \cite{zeegers2023}
and are valuable for understanding the property of interstellar dust.
We compute harmonic IR spectra of these isomers with DFT (PBE0/def2-SVP) and the ML model and show their spectra comparisons in \cref{fig:ir-P5-isomers} (a).
All isomers have MAEs of harmonic frequencies lower than 4 cm$^{-1}$ and infrared intensities are well fitted.
The good transferability of the ML model is expected when the target system is similar with training data in the configuration space even though they may look quite different in the Cartesian coordinate space.
The configuration space for the nanosilicates investigated in this work has high dimensions and is hard to directly visualize.
Therefore, principal component analysis (PCA) is used to reduce the dimension of the configuration space with each configuration represented by a global Smooth Overlap of Atomic Positions (SOAP) feature \cite{SOAP}.
Among the four selected isomers, isomer-I has the lowest MAE of harmonic frequency since it lies closer to training data in the feature space than other isomers in \cref{fig:ir-P5-isomers} (b).
Similarly, isomer-II shows the highest MAE of harmonic frequency because it has the lowest similarity to training data in the feature space.
Later on, transferability analysis is also performed for other pyroxene (P3-P8) and olivine (O2-O7) isomers (see their structure snapshots in Fig. S6 and Fig. S7) which are not directly included in the training data.
The results are shown in Fig. S5.
The MAE of harmonic frequency is 2.1 cm$^{-1}$ for pyroxene and 1.7 cm$^{-1}$ for olivine.
They indicate that the ML model tends to show good prediction accuracy on the test data which looks similar with training data in the feature space.
We do not expect that our ML model will be reliable and accurate for bulk silicates and chemical processes involving dramatic changes in bonding types (e.g. oxygen diffusion, silicate growth and non-stoichiometric silicates).
The transferability to larger cluster sizes will be examined in a separate study.

\subsection{Application of the ML model}
\begin{figure}
	\begin{center}
	\includegraphics[width=\columnwidth]{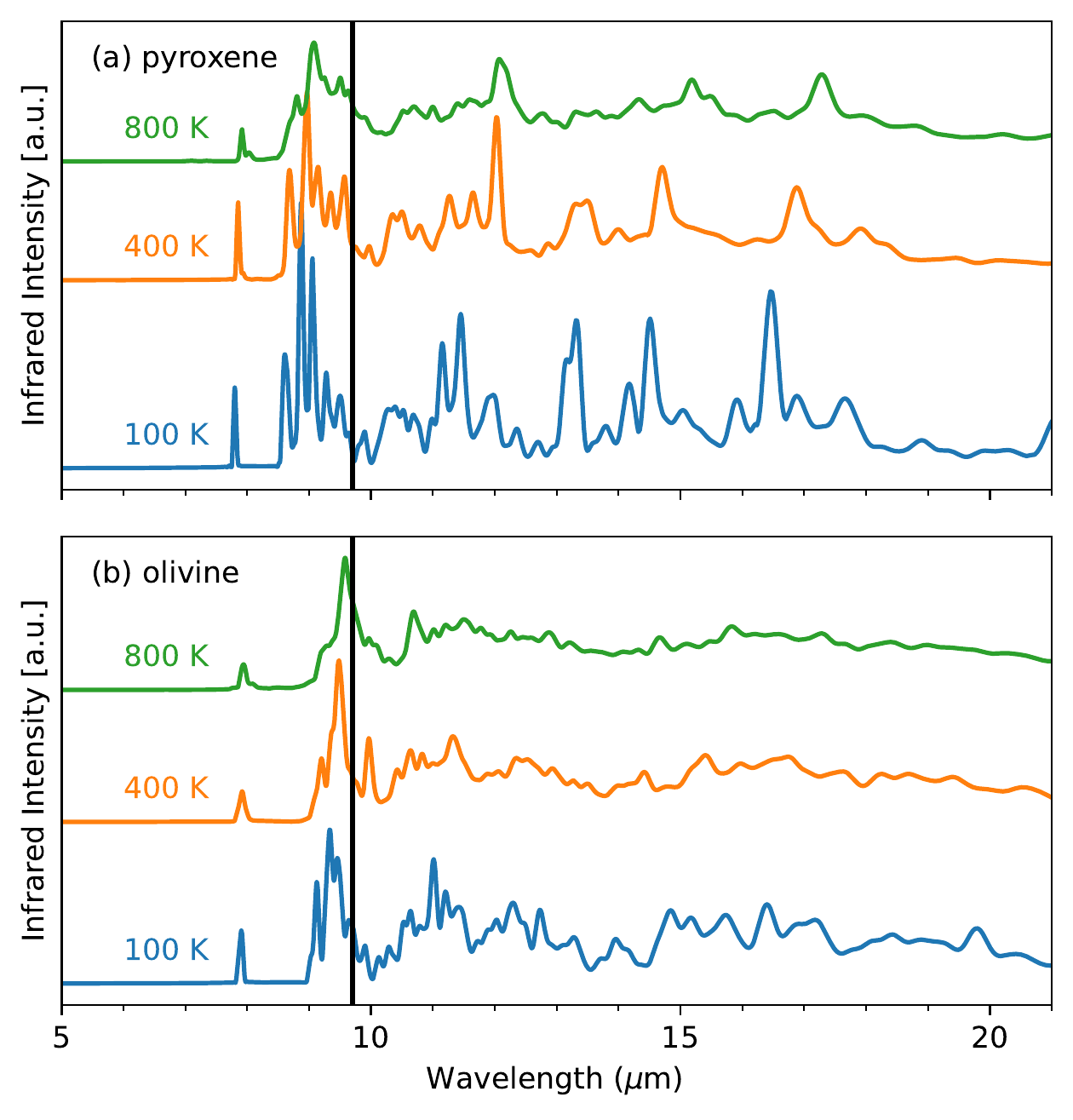}
	\caption{
		Sum of IR spectra of (a) pyroxene (P1-P10) (b) olivine (O1-O10) at 100K, 400K and 800K using the ML model,
		assuming equal weights of different cluster sizes.
		A vertical black line highlights the 9.7 $\mu$m silicate band.}
    \label{fig:ir-sum}
    \end{center}
\end{figure}

As our motivation for developing a ML potential of nanosilicate clusters is to aid the interpretation of silicate band in infrared observations,
we compute IR spectra of 20 silicate clusters (P1-P10 and O1-O10) via MD simulations with the ML model.
\cref{fig:ir-sum} shows the sum of IR spectra for two types (pyroxene and olivine) of silicate clusters while assuming the distribution of cluster size is uniform.
For both pyroxene and olivine clusters, the most intense IR peaks are in the 9-10 $\mu$m range.
They both have a main peak with a slightly smaller wavelength than the signature 9.7 $\mu$m feature at low temperature.
Our MD-based IR spectra at low temperature are in a good agreement with a previous IR study of pyroxene and olivine clusters under the harmonic approximation \cite{escatllar2019}.
As temperature increases, this blueshifted peak broadens and becomes closer to 9.7 $\mu$m. 
In addition, the center position of IR peaks of olivine is closer to 9.7 $\mu$m than that of pyroxene.
%In addition, olivine tends to have a single broad peak in the 9-10 $\mu$m range when the temperature increases, while pyroxene always has multiple peaks regardless of temperature.
%Considering the similarity between the computed IR spectra and the astronomically observed 9.7 $\mu$m band, the dominant existing form of silicate dust should be olivine.
%This agrees well with an estimate of 15.1\% pyroxene and 84.9\% olivine for the amorphous silicate in the diffuse ISM \cite{kemper2004}.
More systematic IR investigations of nanosilicate clusters and their astrophysical implications are planned in the future.

\section{Conclusions}
We have demonstrated that active learning is an efficient method to generate training data for constructing ML interatomic potentials.
After initial training, our ML potential is iteratively improved employing active learning, which can be further accelerated in a parallel manner.
%With this method, a relatively small amount of training data is sufficient to develop an ML potential for nanosilicate clusters with high accuracy respect to DFT-calculated energies, forces and dipoles.
Our comparison of ML-based and DFT-based infrared spectra (\cref{fig:IR-DFT-ML-MD}) reveals that our ML potential is highly accurate for infrared spectroscopic studies of nanosilicate clusters with both harmonic approximation and an MD-based approach.
Our ML potential also exhibits certain transferability regarding harmonic frequencies and IR intensities for high-energy isomers of nanosilicate clusters that are not directly included in the training data, but have similar cluster sizes as the training data.
Since the ML potential is computationally much cheaper than DFT, longer MD runs can be conducted and better statistics can be achieved.
It remains an open question if our ML potential is still accurate and reliable for larger silicate cluster sizes that are not covered in this study.
We will address this issue in a future study.
%We employ our ML model to simulate the IR spectra of silicate clusters with a range of sizes and temperatures.
%We further validate our ML model against harmonic IR spectra of a wide range of silicate clusters and MD-based spectra of selected clusters.
Our work will allow for systemic investigations into how the IR spectra of silicate clusters depend on size, structure and temperature. 
Our work will be particularly beneficial for understanding the properties of silicate dust grains in space, especially in connection with the highly sensitive IR capabilities of the James Webb Space Telescope \cite{zeegers2021}.

\section{Supplementary Material}
See the supplementary material for examples of ORCA input files, silicate structures that are used in this work and more validations of the ML model.

\begin{acknowledgments}
We thank Andreas Møller Slavensky for providing the structure files of high-energy silicate isomers.
This work has been supported by the Danish National Research Foundation through the Center of Excellence “InterCat” (Grant agreement no.: DNRF150) and the VILLUM FONDEN (Investigator grant, Project No. 16562).
S.T.B acknowledges support from the MICINN funded project grants PID2021-127957NB-I00 and TED2021-132550B-C21,
project grant 2021SGR00354 funded by the Generalitat de Catalunya and through the María de Maeztu program for Spanish Structures of Excellence (CEX2021-001202-M). 
\end{acknowledgments}

\section{Conflict of Interest}
The authors have no conflicts to disclose.

\section{DATA AVAILABILITY}
The ML model developed in this work and source data for figures in this article are available in \href{https://github.com/zyt0y/MLP-IR-nanosilicate-clusters}{https://github.com/zyt0y/MLP-IR-nanosilicate-clusters}.
Other data that support the findings of this work are available within the article and its supplementary material.

\section*{References}
\bibliography{ref}

\begin{thebibliography}{88}%
\makeatletter
\providecommand \@ifxundefined [1]{%
 \@ifx{#1\undefined}
}%
\providecommand \@ifnum [1]{%
 \ifnum #1\expandafter \@firstoftwo
 \else \expandafter \@secondoftwo
 \fi
}%
\providecommand \@ifx [1]{%
 \ifx #1\expandafter \@firstoftwo
 \else \expandafter \@secondoftwo
 \fi
}%
\providecommand \natexlab [1]{#1}%
\providecommand \enquote  [1]{``#1''}%
\providecommand \bibnamefont  [1]{#1}%
\providecommand \bibfnamefont [1]{#1}%
\providecommand \citenamefont [1]{#1}%
\providecommand \href@noop [0]{\@secondoftwo}%
\providecommand \href [0]{\begingroup \@sanitize@url \@href}%
\providecommand \@href[1]{\@@startlink{#1}\@@href}%
\providecommand \@@href[1]{\endgroup#1\@@endlink}%
\providecommand \@sanitize@url [0]{\catcode `\\12\catcode `\$12\catcode
  `\&12\catcode `\#12\catcode `\^12\catcode `\_12\catcode `\%12\relax}%
\providecommand \@@startlink[1]{}%
\providecommand \@@endlink[0]{}%
\providecommand \url  [0]{\begingroup\@sanitize@url \@url }%
\providecommand \@url [1]{\endgroup\@href {#1}{\urlprefix }}%
\providecommand \urlprefix  [0]{URL }%
\providecommand \Eprint [0]{\href }%
\providecommand \doibase [0]{http://dx.doi.org/}%
\providecommand \selectlanguage [0]{\@gobble}%
\providecommand \bibinfo  [0]{\@secondoftwo}%
\providecommand \bibfield  [0]{\@secondoftwo}%
\providecommand \translation [1]{[#1]}%
\providecommand \BibitemOpen [0]{}%
\providecommand \bibitemStop [0]{}%
\providecommand \bibitemNoStop [0]{.\EOS\space}%
\providecommand \EOS [0]{\spacefactor3000\relax}%
\providecommand \BibitemShut  [1]{\csname bibitem#1\endcsname}%
\let\auto@bib@innerbib\@empty
%</preamble>
\bibitem [{\citenamefont {Henning}(2010)}]{henning2010}%
  \BibitemOpen
  \bibfield  {author} {\bibinfo {author} {\bibfnamefont {T.}~\bibnamefont
  {Henning}},\ }\href {\doibase 10.1146/annurev-astro-081309-130815} {\bibfield
   {journal} {\bibinfo  {journal} {Annu. Rev. Astron. Astrophys.}\ }\textbf
  {\bibinfo {volume} {48}},\ \bibinfo {pages} {21} (\bibinfo {year}
  {2010})}\BibitemShut {NoStop}%
\bibitem [{\citenamefont {Tielens}\ \emph {et~al.}(1997)\citenamefont
  {Tielens}, \citenamefont {Waters}, \citenamefont {Molster},\ and\
  \citenamefont {Justtanont}}]{tielens1997}%
  \BibitemOpen
  \bibfield  {author} {\bibinfo {author} {\bibfnamefont {A.}~\bibnamefont
  {Tielens}}, \bibinfo {author} {\bibfnamefont {L.}~\bibnamefont {Waters}},
  \bibinfo {author} {\bibfnamefont {F.}~\bibnamefont {Molster}}, \ and\
  \bibinfo {author} {\bibfnamefont {K.}~\bibnamefont {Justtanont}},\ }\href
  {\doibase 10.1023/A:1001585120472} {\bibfield  {journal} {\bibinfo  {journal}
  {Astrophys. Space Sci.}\ }\textbf {\bibinfo {volume} {255}},\ \bibinfo
  {pages} {415} (\bibinfo {year} {1997})}\BibitemShut {NoStop}%
\bibitem [{\citenamefont {Potapov}\ and\ \citenamefont
  {McCoustra}(2021)}]{potapov2021}%
  \BibitemOpen
  \bibfield  {author} {\bibinfo {author} {\bibfnamefont {A.}~\bibnamefont
  {Potapov}}\ and\ \bibinfo {author} {\bibfnamefont {M.}~\bibnamefont
  {McCoustra}},\ }\href {\doibase 10.1080/0144235X.2021.1918498} {\bibfield
  {journal} {\bibinfo  {journal} {Int Rev Phys Chem}\ }\textbf {\bibinfo
  {volume} {40}},\ \bibinfo {pages} {299} (\bibinfo {year} {2021})}\BibitemShut
  {NoStop}%
\bibitem [{\citenamefont {Min}\ \emph {et~al.}(2007)\citenamefont {Min},
  \citenamefont {Waters}, \citenamefont {de~Koter}, \citenamefont {Hovenier},
  \citenamefont {Keller},\ and\ \citenamefont {{Markwick-Kemper}}}]{min2007}%
  \BibitemOpen
  \bibfield  {author} {\bibinfo {author} {\bibfnamefont {M.}~\bibnamefont
  {Min}}, \bibinfo {author} {\bibfnamefont {L.~B. F.~M.}\ \bibnamefont
  {Waters}}, \bibinfo {author} {\bibfnamefont {A.}~\bibnamefont {de~Koter}},
  \bibinfo {author} {\bibfnamefont {J.~W.}\ \bibnamefont {Hovenier}}, \bibinfo
  {author} {\bibfnamefont {L.~P.}\ \bibnamefont {Keller}}, \ and\ \bibinfo
  {author} {\bibfnamefont {F.}~\bibnamefont {{Markwick-Kemper}}},\ }\href
  {\doibase 10.1051/0004-6361:20065436} {\bibfield  {journal} {\bibinfo
  {journal} {A\&A}\ }\textbf {\bibinfo {volume} {462}},\ \bibinfo {pages} {667}
  (\bibinfo {year} {2007})}\BibitemShut {NoStop}%
\bibitem [{\citenamefont {Kemper}, \citenamefont {Vriend},\ and\ \citenamefont
  {Tielens}(2004)}]{kemper2004}%
  \BibitemOpen
  \bibfield  {author} {\bibinfo {author} {\bibfnamefont {F.}~\bibnamefont
  {Kemper}}, \bibinfo {author} {\bibfnamefont {W.~J.}\ \bibnamefont {Vriend}},
  \ and\ \bibinfo {author} {\bibfnamefont {A.~G. G.~M.}\ \bibnamefont
  {Tielens}},\ }\href {\doibase 10.1086/421339} {\bibfield  {journal} {\bibinfo
   {journal} {ApJ}\ }\textbf {\bibinfo {volume} {609}},\ \bibinfo {pages} {826}
  (\bibinfo {year} {2004})}\BibitemShut {NoStop}%
\bibitem [{\citenamefont {Spoon}\ \emph {et~al.}(2022)\citenamefont {Spoon},
  \citenamefont {{Hern{\'a}n-Caballero}}, \citenamefont {Rupke}, \citenamefont
  {Waters}, \citenamefont {Lebouteiller}, \citenamefont {Tielens},
  \citenamefont {Loredo}, \citenamefont {Su},\ and\ \citenamefont
  {Viola}}]{spoon2022}%
  \BibitemOpen
  \bibfield  {author} {\bibinfo {author} {\bibfnamefont {H.~W.~W.}\
  \bibnamefont {Spoon}}, \bibinfo {author} {\bibfnamefont {A.}~\bibnamefont
  {{Hern{\'a}n-Caballero}}}, \bibinfo {author} {\bibfnamefont {D.}~\bibnamefont
  {Rupke}}, \bibinfo {author} {\bibfnamefont {L.~B. F.~M.}\ \bibnamefont
  {Waters}}, \bibinfo {author} {\bibfnamefont {V.}~\bibnamefont
  {Lebouteiller}}, \bibinfo {author} {\bibfnamefont {A.~G. G.~M.}\ \bibnamefont
  {Tielens}}, \bibinfo {author} {\bibfnamefont {T.}~\bibnamefont {Loredo}},
  \bibinfo {author} {\bibfnamefont {Y.}~\bibnamefont {Su}}, \ and\ \bibinfo
  {author} {\bibfnamefont {V.}~\bibnamefont {Viola}},\ }\href {\doibase
  10.3847/1538-4365/ac4989} {\bibfield  {journal} {\bibinfo  {journal} {ApJS}\
  }\textbf {\bibinfo {volume} {259}},\ \bibinfo {pages} {37} (\bibinfo {year}
  {2022})}\BibitemShut {NoStop}%
\bibitem [{\citenamefont {Draine}\ and\ \citenamefont {Li}(2001)}]{draine2001}%
  \BibitemOpen
  \bibfield  {author} {\bibinfo {author} {\bibfnamefont {B.~T.}\ \bibnamefont
  {Draine}}\ and\ \bibinfo {author} {\bibfnamefont {A.}~\bibnamefont {Li}},\
  }\href {\doibase 10.1086/320227} {\bibfield  {journal} {\bibinfo  {journal}
  {ApJ}\ }\textbf {\bibinfo {volume} {551}},\ \bibinfo {pages} {807} (\bibinfo
  {year} {2001})}\BibitemShut {NoStop}%
\bibitem [{\citenamefont {Li}\ and\ \citenamefont
  {Draine}(2001{\natexlab{a}})}]{li2001a}%
  \BibitemOpen
  \bibfield  {author} {\bibinfo {author} {\bibfnamefont {A.}~\bibnamefont
  {Li}}\ and\ \bibinfo {author} {\bibfnamefont {B.~T.}\ \bibnamefont
  {Draine}},\ }\href {\doibase 10.1086/323147} {\bibfield  {journal} {\bibinfo
  {journal} {ApJ}\ }\textbf {\bibinfo {volume} {554}},\ \bibinfo {pages} {778}
  (\bibinfo {year} {2001}{\natexlab{a}})}\BibitemShut {NoStop}%
\bibitem [{\citenamefont {Li}\ and\ \citenamefont
  {Draine}(2001{\natexlab{b}})}]{li2001}%
  \BibitemOpen
  \bibfield  {author} {\bibinfo {author} {\bibfnamefont {A.}~\bibnamefont
  {Li}}\ and\ \bibinfo {author} {\bibfnamefont {B.~T.}\ \bibnamefont
  {Draine}},\ }\href {\doibase 10.1086/319640} {\bibfield  {journal} {\bibinfo
  {journal} {ApJ}\ }\textbf {\bibinfo {volume} {550}},\ \bibinfo {pages} {L213}
  (\bibinfo {year} {2001}{\natexlab{b}})}\BibitemShut {NoStop}%
\bibitem [{\citenamefont {Zamirri}\ \emph {et~al.}(2019)\citenamefont
  {Zamirri}, \citenamefont {Maci{\`a}~Escatllar}, \citenamefont
  {Mari{\~n}oso~Guiu}, \citenamefont {Ugliengo},\ and\ \citenamefont
  {Bromley}}]{zamirri2019}%
  \BibitemOpen
  \bibfield  {author} {\bibinfo {author} {\bibfnamefont {L.}~\bibnamefont
  {Zamirri}}, \bibinfo {author} {\bibfnamefont {A.}~\bibnamefont
  {Maci{\`a}~Escatllar}}, \bibinfo {author} {\bibfnamefont {J.}~\bibnamefont
  {Mari{\~n}oso~Guiu}}, \bibinfo {author} {\bibfnamefont {P.}~\bibnamefont
  {Ugliengo}}, \ and\ \bibinfo {author} {\bibfnamefont {S.~T.}\ \bibnamefont
  {Bromley}},\ }\href {\doibase 10.1021/acsearthspacechem.9b00157} {\bibfield
  {journal} {\bibinfo  {journal} {ACS Earth Space Chem.}\ }\textbf {\bibinfo
  {volume} {3}},\ \bibinfo {pages} {2323} (\bibinfo {year} {2019})}\BibitemShut
  {NoStop}%
\bibitem [{\citenamefont {Escatllar}\ \emph {et~al.}(2019)\citenamefont
  {Escatllar}, \citenamefont {Lazaukas}, \citenamefont {Woodley},\ and\
  \citenamefont {Bromley}}]{escatllar2019}%
  \BibitemOpen
  \bibfield  {author} {\bibinfo {author} {\bibfnamefont {A.~M.}\ \bibnamefont
  {Escatllar}}, \bibinfo {author} {\bibfnamefont {T.}~\bibnamefont {Lazaukas}},
  \bibinfo {author} {\bibfnamefont {S.~M.}\ \bibnamefont {Woodley}}, \ and\
  \bibinfo {author} {\bibfnamefont {S.~T.}\ \bibnamefont {Bromley}},\ }\href
  {\doibase 10.1021/acsearthspacechem.9b00139} {\bibfield  {journal} {\bibinfo
  {journal} {ACS Earth Space Chem.}\ }\textbf {\bibinfo {volume} {3}},\
  \bibinfo {pages} {2390} (\bibinfo {year} {2019})}\BibitemShut {NoStop}%
\bibitem [{\citenamefont {Guiu}, \citenamefont {Escatllar},\ and\ \citenamefont
  {Bromley}(2021)}]{guiu2021}%
  \BibitemOpen
  \bibfield  {author} {\bibinfo {author} {\bibfnamefont {J.~M.}\ \bibnamefont
  {Guiu}}, \bibinfo {author} {\bibfnamefont {A.~M.}\ \bibnamefont {Escatllar}},
  \ and\ \bibinfo {author} {\bibfnamefont {S.~T.}\ \bibnamefont {Bromley}},\
  }\href {\doibase 10.1021/acsearthspacechem.0c00341} {\bibfield  {journal}
  {\bibinfo  {journal} {ACS Earth Space Chem.}\ }\textbf {\bibinfo {volume}
  {5}},\ \bibinfo {pages} {812} (\bibinfo {year} {2021})}\BibitemShut {NoStop}%
\bibitem [{\citenamefont {Sabri}\ \emph {et~al.}(2013)\citenamefont {Sabri},
  \citenamefont {Gavilan}, \citenamefont {J{\"a}ger}, \citenamefont {Lemaire},
  \citenamefont {Vidali}, \citenamefont {Mutschke},\ and\ \citenamefont
  {Henning}}]{sabri2013}%
  \BibitemOpen
  \bibfield  {author} {\bibinfo {author} {\bibfnamefont {T.}~\bibnamefont
  {Sabri}}, \bibinfo {author} {\bibfnamefont {L.}~\bibnamefont {Gavilan}},
  \bibinfo {author} {\bibfnamefont {C.}~\bibnamefont {J{\"a}ger}}, \bibinfo
  {author} {\bibfnamefont {J.~L.}\ \bibnamefont {Lemaire}}, \bibinfo {author}
  {\bibfnamefont {G.}~\bibnamefont {Vidali}}, \bibinfo {author} {\bibfnamefont
  {H.}~\bibnamefont {Mutschke}}, \ and\ \bibinfo {author} {\bibfnamefont
  {T.}~\bibnamefont {Henning}},\ }\href {\doibase 10.1088/0004-637X/780/2/180}
  {\bibfield  {journal} {\bibinfo  {journal} {ApJ}\ }\textbf {\bibinfo {volume}
  {780}},\ \bibinfo {pages} {180} (\bibinfo {year} {2013})}\BibitemShut
  {NoStop}%
\bibitem [{\citenamefont {Mari{\~n}oso~Guiu}\ \emph {et~al.}(2022)\citenamefont
  {Mari{\~n}oso~Guiu}, \citenamefont {Ghejan}, \citenamefont {Bernhardt},
  \citenamefont {Bakker}, \citenamefont {Lang},\ and\ \citenamefont
  {Bromley}}]{marinosoguiu2022}%
  \BibitemOpen
  \bibfield  {author} {\bibinfo {author} {\bibfnamefont {J.}~\bibnamefont
  {Mari{\~n}oso~Guiu}}, \bibinfo {author} {\bibfnamefont {B.-A.}\ \bibnamefont
  {Ghejan}}, \bibinfo {author} {\bibfnamefont {T.~M.}\ \bibnamefont
  {Bernhardt}}, \bibinfo {author} {\bibfnamefont {J.~M.}\ \bibnamefont
  {Bakker}}, \bibinfo {author} {\bibfnamefont {S.~M.}\ \bibnamefont {Lang}}, \
  and\ \bibinfo {author} {\bibfnamefont {S.~T.}\ \bibnamefont {Bromley}},\
  }\href {\doibase 10.1021/acsearthspacechem.2c00186} {\bibfield  {journal}
  {\bibinfo  {journal} {ACS Earth Space Chem.}\ }\textbf {\bibinfo {volume}
  {6}},\ \bibinfo {pages} {2465} (\bibinfo {year} {2022})}\BibitemShut
  {NoStop}%
\bibitem [{\citenamefont {Price}, \citenamefont {Parker},\ and\ \citenamefont
  {Leslie}(1987)}]{price1987}%
  \BibitemOpen
  \bibfield  {author} {\bibinfo {author} {\bibfnamefont {G.~D.}\ \bibnamefont
  {Price}}, \bibinfo {author} {\bibfnamefont {S.~C.}\ \bibnamefont {Parker}}, \
  and\ \bibinfo {author} {\bibfnamefont {M.}~\bibnamefont {Leslie}},\ }\href
  {\doibase 10.1007/BF00308782} {\bibfield  {journal} {\bibinfo  {journal}
  {Phys Chem Minerals}\ }\textbf {\bibinfo {volume} {15}},\ \bibinfo {pages}
  {181} (\bibinfo {year} {1987})}\BibitemShut {NoStop}%
\bibitem [{\citenamefont {Walker}, \citenamefont {Wright},\ and\ \citenamefont
  {Slater}(2003)}]{walker2003}%
  \BibitemOpen
  \bibfield  {author} {\bibinfo {author} {\bibfnamefont {A.~M.}\ \bibnamefont
  {Walker}}, \bibinfo {author} {\bibfnamefont {K.}~\bibnamefont {Wright}}, \
  and\ \bibinfo {author} {\bibfnamefont {B.}~\bibnamefont {Slater}},\ }\href
  {\doibase 10.1007/s00269-003-0358-7} {\bibfield  {journal} {\bibinfo
  {journal} {Phys Chem Minerals}\ }\textbf {\bibinfo {volume} {30}},\ \bibinfo
  {pages} {536} (\bibinfo {year} {2003})}\BibitemShut {NoStop}%
\bibitem [{\citenamefont {Behler}\ and\ \citenamefont
  {Parrinello}(2007)}]{behler2007}%
  \BibitemOpen
  \bibfield  {author} {\bibinfo {author} {\bibfnamefont {J.}~\bibnamefont
  {Behler}}\ and\ \bibinfo {author} {\bibfnamefont {M.}~\bibnamefont
  {Parrinello}},\ }\href {\doibase 10.1103/PhysRevLett.98.146401} {\bibfield
  {journal} {\bibinfo  {journal} {Phys. Rev. Lett.}\ }\textbf {\bibinfo
  {volume} {98}},\ \bibinfo {pages} {146401} (\bibinfo {year}
  {2007})}\BibitemShut {NoStop}%
\bibitem [{\citenamefont {Smith}, \citenamefont {Isayev},\ and\ \citenamefont
  {Roitberg}(2017)}]{ANI-1}%
  \BibitemOpen
  \bibfield  {author} {\bibinfo {author} {\bibfnamefont {J.~S.}\ \bibnamefont
  {Smith}}, \bibinfo {author} {\bibfnamefont {O.}~\bibnamefont {Isayev}}, \
  and\ \bibinfo {author} {\bibfnamefont {A.~E.}\ \bibnamefont {Roitberg}},\
  }\href {\doibase 10.1039/C6SC05720A} {\bibfield  {journal} {\bibinfo
  {journal} {Chem. Sci.}\ }\textbf {\bibinfo {volume} {8}},\ \bibinfo {pages}
  {3192} (\bibinfo {year} {2017})}\BibitemShut {NoStop}%
\bibitem [{\citenamefont {Sch{\"u}tt}\ \emph {et~al.}(2018)\citenamefont
  {Sch{\"u}tt}, \citenamefont {Sauceda}, \citenamefont {Kindermans},
  \citenamefont {Tkatchenko},\ and\ \citenamefont {M{\"u}ller}}]{SchNet}%
  \BibitemOpen
  \bibfield  {author} {\bibinfo {author} {\bibfnamefont {K.~T.}\ \bibnamefont
  {Sch{\"u}tt}}, \bibinfo {author} {\bibfnamefont {H.~E.}\ \bibnamefont
  {Sauceda}}, \bibinfo {author} {\bibfnamefont {P.-J.}\ \bibnamefont
  {Kindermans}}, \bibinfo {author} {\bibfnamefont {A.}~\bibnamefont
  {Tkatchenko}}, \ and\ \bibinfo {author} {\bibfnamefont {K.-R.}\ \bibnamefont
  {M{\"u}ller}},\ }\href {\doibase 10.1063/1.5019779} {\bibfield  {journal}
  {\bibinfo  {journal} {J. Chem. Phys.}\ }\textbf {\bibinfo {volume} {148}},\
  \bibinfo {pages} {241722} (\bibinfo {year} {2018})}\BibitemShut {NoStop}%
\bibitem [{\citenamefont {Unke}\ and\ \citenamefont {Meuwly}(2019)}]{PhysNet}%
  \BibitemOpen
  \bibfield  {author} {\bibinfo {author} {\bibfnamefont {O.~T.}\ \bibnamefont
  {Unke}}\ and\ \bibinfo {author} {\bibfnamefont {M.}~\bibnamefont {Meuwly}},\
  }\href {\doibase 10.1021/acs.jctc.9b00181} {\bibfield  {journal} {\bibinfo
  {journal} {J. Chem. Theory Comput.}\ }\textbf {\bibinfo {volume} {15}},\
  \bibinfo {pages} {3678} (\bibinfo {year} {2019})}\BibitemShut {NoStop}%
\bibitem [{\citenamefont {Zaverkin}\ and\ \citenamefont
  {K{\"a}stner}(2020)}]{GM-NN}%
  \BibitemOpen
  \bibfield  {author} {\bibinfo {author} {\bibfnamefont {V.}~\bibnamefont
  {Zaverkin}}\ and\ \bibinfo {author} {\bibfnamefont {J.}~\bibnamefont
  {K{\"a}stner}},\ }\href {\doibase 10.1021/acs.jctc.0c00347} {\bibfield
  {journal} {\bibinfo  {journal} {J. Chem. Theory Comput.}\ }\textbf {\bibinfo
  {volume} {16}},\ \bibinfo {pages} {5410} (\bibinfo {year}
  {2020})}\BibitemShut {NoStop}%
\bibitem [{\citenamefont {Zaverkin}\ \emph {et~al.}(2021)\citenamefont
  {Zaverkin}, \citenamefont {Holzm{\"u}ller}, \citenamefont {Steinwart},\ and\
  \citenamefont {K{\"a}stner}}]{iGM-NN}%
  \BibitemOpen
  \bibfield  {author} {\bibinfo {author} {\bibfnamefont {V.}~\bibnamefont
  {Zaverkin}}, \bibinfo {author} {\bibfnamefont {D.}~\bibnamefont
  {Holzm{\"u}ller}}, \bibinfo {author} {\bibfnamefont {I.}~\bibnamefont
  {Steinwart}}, \ and\ \bibinfo {author} {\bibfnamefont {J.}~\bibnamefont
  {K{\"a}stner}},\ }\href {\doibase 10.1021/acs.jctc.1c00527} {\bibfield
  {journal} {\bibinfo  {journal} {J. Chem. Theory Comput.}\ }\textbf {\bibinfo
  {volume} {17}},\ \bibinfo {pages} {6658} (\bibinfo {year}
  {2021})}\BibitemShut {NoStop}%
\bibitem [{\citenamefont {Bart{\'o}k}\ \emph {et~al.}(2010)\citenamefont
  {Bart{\'o}k}, \citenamefont {Payne}, \citenamefont {Kondor},\ and\
  \citenamefont {Cs{\'a}nyi}}]{bartok2010}%
  \BibitemOpen
  \bibfield  {author} {\bibinfo {author} {\bibfnamefont {A.~P.}\ \bibnamefont
  {Bart{\'o}k}}, \bibinfo {author} {\bibfnamefont {M.~C.}\ \bibnamefont
  {Payne}}, \bibinfo {author} {\bibfnamefont {R.}~\bibnamefont {Kondor}}, \
  and\ \bibinfo {author} {\bibfnamefont {G.}~\bibnamefont {Cs{\'a}nyi}},\
  }\href {\doibase 10.1103/PhysRevLett.104.136403} {\bibfield  {journal}
  {\bibinfo  {journal} {Phys. Rev. Lett.}\ }\textbf {\bibinfo {volume} {104}},\
  \bibinfo {pages} {136403} (\bibinfo {year} {2010})}\BibitemShut {NoStop}%
\bibitem [{\citenamefont {Koistinen}\ \emph {et~al.}(2017)\citenamefont
  {Koistinen}, \citenamefont {Dagbjartsd{\'o}ttir}, \citenamefont
  {{\'A}sgeirsson}, \citenamefont {Vehtari},\ and\ \citenamefont
  {J{\'o}nsson}}]{koistinen2017}%
  \BibitemOpen
  \bibfield  {author} {\bibinfo {author} {\bibfnamefont {O.-P.}\ \bibnamefont
  {Koistinen}}, \bibinfo {author} {\bibfnamefont {F.~B.}\ \bibnamefont
  {Dagbjartsd{\'o}ttir}}, \bibinfo {author} {\bibfnamefont {V.}~\bibnamefont
  {{\'A}sgeirsson}}, \bibinfo {author} {\bibfnamefont {A.}~\bibnamefont
  {Vehtari}}, \ and\ \bibinfo {author} {\bibfnamefont {H.}~\bibnamefont
  {J{\'o}nsson}},\ }\href {\doibase 10.1063/1.4986787} {\bibfield  {journal}
  {\bibinfo  {journal} {J. Chem. Phys.}\ }\textbf {\bibinfo {volume} {147}},\
  \bibinfo {pages} {152720} (\bibinfo {year} {2017})}\BibitemShut {NoStop}%
\bibitem [{\citenamefont {Denzel}\ and\ \citenamefont
  {K{\"a}stner}(2018)}]{denzel2018a}%
  \BibitemOpen
  \bibfield  {author} {\bibinfo {author} {\bibfnamefont {A.}~\bibnamefont
  {Denzel}}\ and\ \bibinfo {author} {\bibfnamefont {J.}~\bibnamefont
  {K{\"a}stner}},\ }\href {\doibase 10.1063/1.5017103} {\bibfield  {journal}
  {\bibinfo  {journal} {J. Chem. Phys.}\ }\textbf {\bibinfo {volume} {148}},\
  \bibinfo {pages} {094114} (\bibinfo {year} {2018})}\BibitemShut {NoStop}%
\bibitem [{\citenamefont {Behler}(2021)}]{behler2021}%
  \BibitemOpen
  \bibfield  {author} {\bibinfo {author} {\bibfnamefont {J.}~\bibnamefont
  {Behler}},\ }\href {\doibase 10.1021/acs.chemrev.0c00868} {\bibfield
  {journal} {\bibinfo  {journal} {Chem. Rev.}\ }\textbf {\bibinfo {volume}
  {121}},\ \bibinfo {pages} {10037} (\bibinfo {year} {2021})}\BibitemShut
  {NoStop}%
\bibitem [{\citenamefont {Deringer}\ \emph
  {et~al.}(2021{\natexlab{a}})\citenamefont {Deringer}, \citenamefont
  {Bart{\'o}k}, \citenamefont {Bernstein}, \citenamefont {Wilkins},
  \citenamefont {Ceriotti},\ and\ \citenamefont {Cs{\'a}nyi}}]{deringer2021a}%
  \BibitemOpen
  \bibfield  {author} {\bibinfo {author} {\bibfnamefont {V.~L.}\ \bibnamefont
  {Deringer}}, \bibinfo {author} {\bibfnamefont {A.~P.}\ \bibnamefont
  {Bart{\'o}k}}, \bibinfo {author} {\bibfnamefont {N.}~\bibnamefont
  {Bernstein}}, \bibinfo {author} {\bibfnamefont {D.~M.}\ \bibnamefont
  {Wilkins}}, \bibinfo {author} {\bibfnamefont {M.}~\bibnamefont {Ceriotti}}, \
  and\ \bibinfo {author} {\bibfnamefont {G.}~\bibnamefont {Cs{\'a}nyi}},\
  }\href {\doibase 10.1021/acs.chemrev.1c00022} {\bibfield  {journal} {\bibinfo
   {journal} {Chem. Rev.}\ }\textbf {\bibinfo {volume} {121}},\ \bibinfo
  {pages} {10073} (\bibinfo {year} {2021}{\natexlab{a}})}\BibitemShut {NoStop}%
\bibitem [{\citenamefont {Ouyang}, \citenamefont {Xie},\ and\ \citenamefont
  {Jiang}(2015)}]{ouyang2015}%
  \BibitemOpen
  \bibfield  {author} {\bibinfo {author} {\bibfnamefont {R.}~\bibnamefont
  {Ouyang}}, \bibinfo {author} {\bibfnamefont {Y.}~\bibnamefont {Xie}}, \ and\
  \bibinfo {author} {\bibfnamefont {D.-e.}\ \bibnamefont {Jiang}},\ }\href
  {\doibase 10.1039/C5NR03903G} {\bibfield  {journal} {\bibinfo  {journal}
  {Nanoscale}\ }\textbf {\bibinfo {volume} {7}},\ \bibinfo {pages} {14817}
  (\bibinfo {year} {2015})}\BibitemShut {NoStop}%
\bibitem [{\citenamefont {Chiriki}, \citenamefont {Jindal},\ and\ \citenamefont
  {Bulusu}(2017)}]{chiriki2017}%
  \BibitemOpen
  \bibfield  {author} {\bibinfo {author} {\bibfnamefont {S.}~\bibnamefont
  {Chiriki}}, \bibinfo {author} {\bibfnamefont {S.}~\bibnamefont {Jindal}}, \
  and\ \bibinfo {author} {\bibfnamefont {S.~S.}\ \bibnamefont {Bulusu}},\
  }\href {\doibase 10.1063/1.4977050} {\bibfield  {journal} {\bibinfo
  {journal} {J. Chem. Phys.}\ }\textbf {\bibinfo {volume} {146}},\ \bibinfo
  {pages} {084314} (\bibinfo {year} {2017})}\BibitemShut {NoStop}%
\bibitem [{\citenamefont {Bisbo}\ and\ \citenamefont {Hammer}(2020)}]{GOFEE}%
  \BibitemOpen
  \bibfield  {author} {\bibinfo {author} {\bibfnamefont {M.~K.}\ \bibnamefont
  {Bisbo}}\ and\ \bibinfo {author} {\bibfnamefont {B.}~\bibnamefont {Hammer}},\
  }\href {\doibase 10.1103/PhysRevLett.124.086102} {\bibfield  {journal}
  {\bibinfo  {journal} {Phys. Rev. Lett.}\ }\textbf {\bibinfo {volume} {124}},\
  \bibinfo {pages} {086102} (\bibinfo {year} {2020})}\BibitemShut {NoStop}%
\bibitem [{\citenamefont {Paleico}\ and\ \citenamefont
  {Behler}(2020{\natexlab{a}})}]{paleico2020}%
  \BibitemOpen
  \bibfield  {author} {\bibinfo {author} {\bibfnamefont {M.~L.}\ \bibnamefont
  {Paleico}}\ and\ \bibinfo {author} {\bibfnamefont {J.}~\bibnamefont
  {Behler}},\ }\href {\doibase 10.1063/1.5142363} {\bibfield  {journal}
  {\bibinfo  {journal} {J. Chem. Phys.}\ }\textbf {\bibinfo {volume} {152}},\
  \bibinfo {pages} {094109} (\bibinfo {year} {2020}{\natexlab{a}})}\BibitemShut
  {NoStop}%
\bibitem [{\citenamefont {Paleico}\ and\ \citenamefont
  {Behler}(2020{\natexlab{b}})}]{paleico2020a}%
  \BibitemOpen
  \bibfield  {author} {\bibinfo {author} {\bibfnamefont {M.~L.}\ \bibnamefont
  {Paleico}}\ and\ \bibinfo {author} {\bibfnamefont {J.}~\bibnamefont
  {Behler}},\ }\href {\doibase 10.1063/5.0014876} {\bibfield  {journal}
  {\bibinfo  {journal} {J. Chem. Phys.}\ }\textbf {\bibinfo {volume} {153}},\
  \bibinfo {pages} {054704} (\bibinfo {year} {2020}{\natexlab{b}})}\BibitemShut
  {NoStop}%
\bibitem [{\citenamefont {Timmermann}\ \emph {et~al.}(2020)\citenamefont
  {Timmermann}, \citenamefont {Kraushofer}, \citenamefont {Resch},
  \citenamefont {Li}, \citenamefont {Wang}, \citenamefont {Mao}, \citenamefont
  {Riva}, \citenamefont {Lee}, \citenamefont {Staacke}, \citenamefont {Schmid},
  \citenamefont {Scheurer}, \citenamefont {Parkinson}, \citenamefont
  {Diebold},\ and\ \citenamefont {Reuter}}]{timmermann2020}%
  \BibitemOpen
  \bibfield  {author} {\bibinfo {author} {\bibfnamefont {J.}~\bibnamefont
  {Timmermann}}, \bibinfo {author} {\bibfnamefont {F.}~\bibnamefont
  {Kraushofer}}, \bibinfo {author} {\bibfnamefont {N.}~\bibnamefont {Resch}},
  \bibinfo {author} {\bibfnamefont {P.}~\bibnamefont {Li}}, \bibinfo {author}
  {\bibfnamefont {Y.}~\bibnamefont {Wang}}, \bibinfo {author} {\bibfnamefont
  {Z.}~\bibnamefont {Mao}}, \bibinfo {author} {\bibfnamefont {M.}~\bibnamefont
  {Riva}}, \bibinfo {author} {\bibfnamefont {Y.}~\bibnamefont {Lee}}, \bibinfo
  {author} {\bibfnamefont {C.}~\bibnamefont {Staacke}}, \bibinfo {author}
  {\bibfnamefont {M.}~\bibnamefont {Schmid}}, \bibinfo {author} {\bibfnamefont
  {C.}~\bibnamefont {Scheurer}}, \bibinfo {author} {\bibfnamefont {G.~S.}\
  \bibnamefont {Parkinson}}, \bibinfo {author} {\bibfnamefont {U.}~\bibnamefont
  {Diebold}}, \ and\ \bibinfo {author} {\bibfnamefont {K.}~\bibnamefont
  {Reuter}},\ }\href {\doibase 10.1103/PhysRevLett.125.206101} {\bibfield
  {journal} {\bibinfo  {journal} {Phys. Rev. Lett.}\ }\textbf {\bibinfo
  {volume} {125}},\ \bibinfo {pages} {206101} (\bibinfo {year}
  {2020})}\BibitemShut {NoStop}%
\bibitem [{\citenamefont {Kaappa}, \citenamefont {{del R{\'i}o}},\ and\
  \citenamefont {Jacobsen}(2021)}]{kaappa2021}%
  \BibitemOpen
  \bibfield  {author} {\bibinfo {author} {\bibfnamefont {S.}~\bibnamefont
  {Kaappa}}, \bibinfo {author} {\bibfnamefont {E.~G.}\ \bibnamefont {{del
  R{\'i}o}}}, \ and\ \bibinfo {author} {\bibfnamefont {K.~W.}\ \bibnamefont
  {Jacobsen}},\ }\href {\doibase 10.1103/PhysRevB.103.174114} {\bibfield
  {journal} {\bibinfo  {journal} {Phys. Rev. B}\ }\textbf {\bibinfo {volume}
  {103}},\ \bibinfo {pages} {174114} (\bibinfo {year} {2021})}\BibitemShut
  {NoStop}%
\bibitem [{\citenamefont {Bisbo}\ and\ \citenamefont
  {Hammer}(2022)}]{bisbo2022}%
  \BibitemOpen
  \bibfield  {author} {\bibinfo {author} {\bibfnamefont {M.~K.}\ \bibnamefont
  {Bisbo}}\ and\ \bibinfo {author} {\bibfnamefont {B.}~\bibnamefont {Hammer}},\
  }\href {\doibase 10.1103/PhysRevB.105.245404} {\bibfield  {journal} {\bibinfo
   {journal} {Phys. Rev. B}\ }\textbf {\bibinfo {volume} {105}},\ \bibinfo
  {pages} {245404} (\bibinfo {year} {2022})}\BibitemShut {NoStop}%
\bibitem [{\citenamefont {Christiansen}, \citenamefont {R{\o}nne},\ and\
  \citenamefont {Hammer}(2022)}]{christiansen2022}%
  \BibitemOpen
  \bibfield  {author} {\bibinfo {author} {\bibfnamefont {M.-P.~V.}\
  \bibnamefont {Christiansen}}, \bibinfo {author} {\bibfnamefont
  {N.}~\bibnamefont {R{\o}nne}}, \ and\ \bibinfo {author} {\bibfnamefont
  {B.}~\bibnamefont {Hammer}},\ }\href {\doibase 10.1063/5.0094165} {\bibfield
  {journal} {\bibinfo  {journal} {J. Chem. Phys.}\ }\textbf {\bibinfo {volume}
  {157}},\ \bibinfo {pages} {054701} (\bibinfo {year} {2022})}\BibitemShut
  {NoStop}%
\bibitem [{\citenamefont {R{\o}nne}\ \emph {et~al.}(2022)\citenamefont
  {R{\o}nne}, \citenamefont {Christiansen}, \citenamefont {Slavensky},
  \citenamefont {Tang}, \citenamefont {Brix}, \citenamefont {Pedersen},
  \citenamefont {Bisbo},\ and\ \citenamefont {Hammer}}]{ronne2022}%
  \BibitemOpen
  \bibfield  {author} {\bibinfo {author} {\bibfnamefont {N.}~\bibnamefont
  {R{\o}nne}}, \bibinfo {author} {\bibfnamefont {M.-P.~V.}\ \bibnamefont
  {Christiansen}}, \bibinfo {author} {\bibfnamefont {A.~M.}\ \bibnamefont
  {Slavensky}}, \bibinfo {author} {\bibfnamefont {Z.}~\bibnamefont {Tang}},
  \bibinfo {author} {\bibfnamefont {F.}~\bibnamefont {Brix}}, \bibinfo {author}
  {\bibfnamefont {M.~E.}\ \bibnamefont {Pedersen}}, \bibinfo {author}
  {\bibfnamefont {M.~K.}\ \bibnamefont {Bisbo}}, \ and\ \bibinfo {author}
  {\bibfnamefont {B.}~\bibnamefont {Hammer}},\ }\href {\doibase
  10.1063/5.0121748} {\bibfield  {journal} {\bibinfo  {journal} {J. Chem.
  Phys.}\ }\textbf {\bibinfo {volume} {157}},\ \bibinfo {pages} {174115}
  (\bibinfo {year} {2022})}\BibitemShut {NoStop}%
\bibitem [{\citenamefont {Li}, \citenamefont {Kermode},\ and\ \citenamefont
  {De~Vita}(2015)}]{li2015d}%
  \BibitemOpen
  \bibfield  {author} {\bibinfo {author} {\bibfnamefont {Z.}~\bibnamefont
  {Li}}, \bibinfo {author} {\bibfnamefont {J.~R.}\ \bibnamefont {Kermode}}, \
  and\ \bibinfo {author} {\bibfnamefont {A.}~\bibnamefont {De~Vita}},\ }\href
  {\doibase 10.1103/PhysRevLett.114.096405} {\bibfield  {journal} {\bibinfo
  {journal} {Phys. Rev. Lett.}\ }\textbf {\bibinfo {volume} {114}},\ \bibinfo
  {pages} {096405} (\bibinfo {year} {2015})}\BibitemShut {NoStop}%
\bibitem [{\citenamefont {Caro}\ \emph {et~al.}(2018)\citenamefont {Caro},
  \citenamefont {Deringer}, \citenamefont {Koskinen}, \citenamefont {Laurila},\
  and\ \citenamefont {Cs{\'a}nyi}}]{caro2018}%
  \BibitemOpen
  \bibfield  {author} {\bibinfo {author} {\bibfnamefont {M.~A.}\ \bibnamefont
  {Caro}}, \bibinfo {author} {\bibfnamefont {V.~L.}\ \bibnamefont {Deringer}},
  \bibinfo {author} {\bibfnamefont {J.}~\bibnamefont {Koskinen}}, \bibinfo
  {author} {\bibfnamefont {T.}~\bibnamefont {Laurila}}, \ and\ \bibinfo
  {author} {\bibfnamefont {G.}~\bibnamefont {Cs{\'a}nyi}},\ }\href {\doibase
  10.1103/PhysRevLett.120.166101} {\bibfield  {journal} {\bibinfo  {journal}
  {Phys. Rev. Lett.}\ }\textbf {\bibinfo {volume} {120}},\ \bibinfo {pages}
  {166101} (\bibinfo {year} {2018})}\BibitemShut {NoStop}%
\bibitem [{\citenamefont {Caro}\ \emph {et~al.}(2020)\citenamefont {Caro},
  \citenamefont {Cs{\'a}nyi}, \citenamefont {Laurila},\ and\ \citenamefont
  {Deringer}}]{caro2020}%
  \BibitemOpen
  \bibfield  {author} {\bibinfo {author} {\bibfnamefont {M.~A.}\ \bibnamefont
  {Caro}}, \bibinfo {author} {\bibfnamefont {G.}~\bibnamefont {Cs{\'a}nyi}},
  \bibinfo {author} {\bibfnamefont {T.}~\bibnamefont {Laurila}}, \ and\
  \bibinfo {author} {\bibfnamefont {V.~L.}\ \bibnamefont {Deringer}},\ }\href
  {\doibase 10.1103/PhysRevB.102.174201} {\bibfield  {journal} {\bibinfo
  {journal} {Phys. Rev. B}\ }\textbf {\bibinfo {volume} {102}},\ \bibinfo
  {pages} {174201} (\bibinfo {year} {2020})}\BibitemShut {NoStop}%
\bibitem [{\citenamefont {Lim}\ \emph {et~al.}(2020)\citenamefont {Lim},
  \citenamefont {Vandermause}, \citenamefont {{van Spronsen}}, \citenamefont
  {Musaelian}, \citenamefont {Xie}, \citenamefont {Sun}, \citenamefont
  {O'Connor}, \citenamefont {Egle}, \citenamefont {Molinari}, \citenamefont
  {Florian}, \citenamefont {Duanmu}, \citenamefont {Madix}, \citenamefont
  {Sautet}, \citenamefont {Friend},\ and\ \citenamefont {Kozinsky}}]{lim2020}%
  \BibitemOpen
  \bibfield  {author} {\bibinfo {author} {\bibfnamefont {J.~S.}\ \bibnamefont
  {Lim}}, \bibinfo {author} {\bibfnamefont {J.}~\bibnamefont {Vandermause}},
  \bibinfo {author} {\bibfnamefont {M.~A.}\ \bibnamefont {{van Spronsen}}},
  \bibinfo {author} {\bibfnamefont {A.}~\bibnamefont {Musaelian}}, \bibinfo
  {author} {\bibfnamefont {Y.}~\bibnamefont {Xie}}, \bibinfo {author}
  {\bibfnamefont {L.}~\bibnamefont {Sun}}, \bibinfo {author} {\bibfnamefont
  {C.~R.}\ \bibnamefont {O'Connor}}, \bibinfo {author} {\bibfnamefont
  {T.}~\bibnamefont {Egle}}, \bibinfo {author} {\bibfnamefont {N.}~\bibnamefont
  {Molinari}}, \bibinfo {author} {\bibfnamefont {J.}~\bibnamefont {Florian}},
  \bibinfo {author} {\bibfnamefont {K.}~\bibnamefont {Duanmu}}, \bibinfo
  {author} {\bibfnamefont {R.~J.}\ \bibnamefont {Madix}}, \bibinfo {author}
  {\bibfnamefont {P.}~\bibnamefont {Sautet}}, \bibinfo {author} {\bibfnamefont
  {C.~M.}\ \bibnamefont {Friend}}, \ and\ \bibinfo {author} {\bibfnamefont
  {B.}~\bibnamefont {Kozinsky}},\ }\href {\doibase 10.1021/jacs.0c06401}
  {\bibfield  {journal} {\bibinfo  {journal} {J. Am. Chem. Soc.}\ }\textbf
  {\bibinfo {volume} {142}},\ \bibinfo {pages} {15907} (\bibinfo {year}
  {2020})}\BibitemShut {NoStop}%
\bibitem [{\citenamefont {No{\'e}}\ \emph {et~al.}(2020)\citenamefont
  {No{\'e}}, \citenamefont {Tkatchenko}, \citenamefont {M{\"u}ller},\ and\
  \citenamefont {Clementi}}]{noe2020}%
  \BibitemOpen
  \bibfield  {author} {\bibinfo {author} {\bibfnamefont {F.}~\bibnamefont
  {No{\'e}}}, \bibinfo {author} {\bibfnamefont {A.}~\bibnamefont {Tkatchenko}},
  \bibinfo {author} {\bibfnamefont {K.-R.}\ \bibnamefont {M{\"u}ller}}, \ and\
  \bibinfo {author} {\bibfnamefont {C.}~\bibnamefont {Clementi}},\ }\href
  {\doibase 10.1146/annurev-physchem-042018-052331} {\bibfield  {journal}
  {\bibinfo  {journal} {Annu. Rev. Phys. Chem.}\ }\textbf {\bibinfo {volume}
  {71}},\ \bibinfo {pages} {361} (\bibinfo {year} {2020})}\BibitemShut
  {NoStop}%
\bibitem [{\citenamefont {Deringer}\ \emph
  {et~al.}(2021{\natexlab{b}})\citenamefont {Deringer}, \citenamefont
  {Bernstein}, \citenamefont {Cs{\'a}nyi}, \citenamefont {Ben~Mahmoud},
  \citenamefont {Ceriotti}, \citenamefont {Wilson}, \citenamefont {Drabold},\
  and\ \citenamefont {Elliott}}]{deringer2021}%
  \BibitemOpen
  \bibfield  {author} {\bibinfo {author} {\bibfnamefont {V.~L.}\ \bibnamefont
  {Deringer}}, \bibinfo {author} {\bibfnamefont {N.}~\bibnamefont {Bernstein}},
  \bibinfo {author} {\bibfnamefont {G.}~\bibnamefont {Cs{\'a}nyi}}, \bibinfo
  {author} {\bibfnamefont {C.}~\bibnamefont {Ben~Mahmoud}}, \bibinfo {author}
  {\bibfnamefont {M.}~\bibnamefont {Ceriotti}}, \bibinfo {author}
  {\bibfnamefont {M.}~\bibnamefont {Wilson}}, \bibinfo {author} {\bibfnamefont
  {D.~A.}\ \bibnamefont {Drabold}}, \ and\ \bibinfo {author} {\bibfnamefont
  {S.~R.}\ \bibnamefont {Elliott}},\ }\href {\doibase
  10.1038/s41586-020-03072-z} {\bibfield  {journal} {\bibinfo  {journal}
  {Nature}\ }\textbf {\bibinfo {volume} {589}},\ \bibinfo {pages} {59}
  (\bibinfo {year} {2021}{\natexlab{b}})}\BibitemShut {NoStop}%
\bibitem [{\citenamefont {Westermayr}\ \emph {et~al.}(2021)\citenamefont
  {Westermayr}, \citenamefont {Gastegger}, \citenamefont {Sch{\"u}tt},\ and\
  \citenamefont {Maurer}}]{westermayr2021}%
  \BibitemOpen
  \bibfield  {author} {\bibinfo {author} {\bibfnamefont {J.}~\bibnamefont
  {Westermayr}}, \bibinfo {author} {\bibfnamefont {M.}~\bibnamefont
  {Gastegger}}, \bibinfo {author} {\bibfnamefont {K.~T.}\ \bibnamefont
  {Sch{\"u}tt}}, \ and\ \bibinfo {author} {\bibfnamefont {R.~J.}\ \bibnamefont
  {Maurer}},\ }\href {\doibase 10.1063/5.0047760} {\bibfield  {journal}
  {\bibinfo  {journal} {J. Chem. Phys.}\ }\textbf {\bibinfo {volume} {154}},\
  \bibinfo {pages} {230903} (\bibinfo {year} {2021})}\BibitemShut {NoStop}%
\bibitem [{\citenamefont {Unke}\ \emph {et~al.}(2021)\citenamefont {Unke},
  \citenamefont {Chmiela}, \citenamefont {Sauceda}, \citenamefont {Gastegger},
  \citenamefont {Poltavsky}, \citenamefont {Sch{\"u}tt}, \citenamefont
  {Tkatchenko},\ and\ \citenamefont {M{\"u}ller}}]{unke2021}%
  \BibitemOpen
  \bibfield  {author} {\bibinfo {author} {\bibfnamefont {O.~T.}\ \bibnamefont
  {Unke}}, \bibinfo {author} {\bibfnamefont {S.}~\bibnamefont {Chmiela}},
  \bibinfo {author} {\bibfnamefont {H.~E.}\ \bibnamefont {Sauceda}}, \bibinfo
  {author} {\bibfnamefont {M.}~\bibnamefont {Gastegger}}, \bibinfo {author}
  {\bibfnamefont {I.}~\bibnamefont {Poltavsky}}, \bibinfo {author}
  {\bibfnamefont {K.~T.}\ \bibnamefont {Sch{\"u}tt}}, \bibinfo {author}
  {\bibfnamefont {A.}~\bibnamefont {Tkatchenko}}, \ and\ \bibinfo {author}
  {\bibfnamefont {K.-R.}\ \bibnamefont {M{\"u}ller}},\ }\href {\doibase
  10.1021/acs.chemrev.0c01111} {\bibfield  {journal} {\bibinfo  {journal}
  {Chem. Rev.}\ }\textbf {\bibinfo {volume} {121}},\ \bibinfo {pages} {10142}
  (\bibinfo {year} {2021})}\BibitemShut {NoStop}%
\bibitem [{\citenamefont {Musaelian}\ \emph {et~al.}(2023)\citenamefont
  {Musaelian}, \citenamefont {Batzner}, \citenamefont {Johansson},
  \citenamefont {Sun}, \citenamefont {Owen}, \citenamefont {Kornbluth},\ and\
  \citenamefont {Kozinsky}}]{musaelian2023}%
  \BibitemOpen
  \bibfield  {author} {\bibinfo {author} {\bibfnamefont {A.}~\bibnamefont
  {Musaelian}}, \bibinfo {author} {\bibfnamefont {S.}~\bibnamefont {Batzner}},
  \bibinfo {author} {\bibfnamefont {A.}~\bibnamefont {Johansson}}, \bibinfo
  {author} {\bibfnamefont {L.}~\bibnamefont {Sun}}, \bibinfo {author}
  {\bibfnamefont {C.~J.}\ \bibnamefont {Owen}}, \bibinfo {author}
  {\bibfnamefont {M.}~\bibnamefont {Kornbluth}}, \ and\ \bibinfo {author}
  {\bibfnamefont {B.}~\bibnamefont {Kozinsky}},\ }\href {\doibase
  10.1038/s41467-023-36329-y} {\bibfield  {journal} {\bibinfo  {journal} {Nat
  Commun}\ }\textbf {\bibinfo {volume} {14}},\ \bibinfo {pages} {579} (\bibinfo
  {year} {2023})}\BibitemShut {NoStop}%
\bibitem [{\citenamefont {Gastegger}, \citenamefont {Behler},\ and\
  \citenamefont {Marquetand}(2017)}]{gastegger2017}%
  \BibitemOpen
  \bibfield  {author} {\bibinfo {author} {\bibfnamefont {M.}~\bibnamefont
  {Gastegger}}, \bibinfo {author} {\bibfnamefont {J.}~\bibnamefont {Behler}}, \
  and\ \bibinfo {author} {\bibfnamefont {P.}~\bibnamefont {Marquetand}},\
  }\href {\doibase 10.1039/C7SC02267K} {\bibfield  {journal} {\bibinfo
  {journal} {Chem. Sci.}\ }\textbf {\bibinfo {volume} {8}},\ \bibinfo {pages}
  {6924} (\bibinfo {year} {2017})}\BibitemShut {NoStop}%
\bibitem [{\citenamefont {Lam}, \citenamefont {{Abdul-Al}},\ and\ \citenamefont
  {Allouche}(2020)}]{lam2020}%
  \BibitemOpen
  \bibfield  {author} {\bibinfo {author} {\bibfnamefont {J.}~\bibnamefont
  {Lam}}, \bibinfo {author} {\bibfnamefont {S.}~\bibnamefont {{Abdul-Al}}}, \
  and\ \bibinfo {author} {\bibfnamefont {A.-R.}\ \bibnamefont {Allouche}},\
  }\href {\doibase 10.1021/acs.jctc.9b00964} {\bibfield  {journal} {\bibinfo
  {journal} {J. Chem. Theory Comput.}\ }\textbf {\bibinfo {volume} {16}},\
  \bibinfo {pages} {1681} (\bibinfo {year} {2020})}\BibitemShut {NoStop}%
\bibitem [{\citenamefont {Gastegger}, \citenamefont {T.~Sch{\"u}tt},\ and\
  \citenamefont {M{\"u}ller}(2021)}]{FieldSchNet}%
  \BibitemOpen
  \bibfield  {author} {\bibinfo {author} {\bibfnamefont {M.}~\bibnamefont
  {Gastegger}}, \bibinfo {author} {\bibfnamefont {K.}~\bibnamefont
  {T.~Sch{\"u}tt}}, \ and\ \bibinfo {author} {\bibfnamefont {K.-R.}\
  \bibnamefont {M{\"u}ller}},\ }\href {\doibase 10.1039/D1SC02742E} {\bibfield
  {journal} {\bibinfo  {journal} {Chem. Sci.}\ }\textbf {\bibinfo {volume}
  {12}},\ \bibinfo {pages} {11473} (\bibinfo {year} {2021})}\BibitemShut
  {NoStop}%
\bibitem [{\citenamefont {K{\"a}ser}\ \emph {et~al.}(2021)\citenamefont
  {K{\"a}ser}, \citenamefont {Boittier}, \citenamefont {Upadhyay},\ and\
  \citenamefont {Meuwly}}]{kaser2021}%
  \BibitemOpen
  \bibfield  {author} {\bibinfo {author} {\bibfnamefont {S.}~\bibnamefont
  {K{\"a}ser}}, \bibinfo {author} {\bibfnamefont {E.~D.}\ \bibnamefont
  {Boittier}}, \bibinfo {author} {\bibfnamefont {M.}~\bibnamefont {Upadhyay}},
  \ and\ \bibinfo {author} {\bibfnamefont {M.}~\bibnamefont {Meuwly}},\ }\href
  {\doibase 10.1021/acs.jctc.1c00249} {\bibfield  {journal} {\bibinfo
  {journal} {J. Chem. Theory Comput.}\ }\textbf {\bibinfo {volume} {17}},\
  \bibinfo {pages} {3687} (\bibinfo {year} {2021})}\BibitemShut {NoStop}%
\bibitem [{\citenamefont {Beckmann}\ \emph {et~al.}(2022)\citenamefont
  {Beckmann}, \citenamefont {Brieuc}, \citenamefont {Schran},\ and\
  \citenamefont {Marx}}]{beckmann2022}%
  \BibitemOpen
  \bibfield  {author} {\bibinfo {author} {\bibfnamefont {R.}~\bibnamefont
  {Beckmann}}, \bibinfo {author} {\bibfnamefont {F.}~\bibnamefont {Brieuc}},
  \bibinfo {author} {\bibfnamefont {C.}~\bibnamefont {Schran}}, \ and\ \bibinfo
  {author} {\bibfnamefont {D.}~\bibnamefont {Marx}},\ }\href {\doibase
  10.1021/acs.jctc.2c00511} {\bibfield  {journal} {\bibinfo  {journal} {J.
  Chem. Theory Comput.}\ }\textbf {\bibinfo {volume} {18}},\ \bibinfo {pages}
  {5492} (\bibinfo {year} {2022})}\BibitemShut {NoStop}%
\bibitem [{\citenamefont {Le}, \citenamefont {Huynh},\ and\ \citenamefont
  {Raff}(2009)}]{le2009}%
  \BibitemOpen
  \bibfield  {author} {\bibinfo {author} {\bibfnamefont {H.~M.}\ \bibnamefont
  {Le}}, \bibinfo {author} {\bibfnamefont {S.}~\bibnamefont {Huynh}}, \ and\
  \bibinfo {author} {\bibfnamefont {L.~M.}\ \bibnamefont {Raff}},\ }\href
  {\doibase 10.1063/1.3159748} {\bibfield  {journal} {\bibinfo  {journal} {J.
  Chem. Phys.}\ }\textbf {\bibinfo {volume} {131}},\ \bibinfo {pages} {014107}
  (\bibinfo {year} {2009})}\BibitemShut {NoStop}%
\bibitem [{\citenamefont {Artrith}\ and\ \citenamefont
  {Behler}(2012)}]{artrith2012}%
  \BibitemOpen
  \bibfield  {author} {\bibinfo {author} {\bibfnamefont {N.}~\bibnamefont
  {Artrith}}\ and\ \bibinfo {author} {\bibfnamefont {J.}~\bibnamefont
  {Behler}},\ }\href {\doibase 10.1103/PhysRevB.85.045439} {\bibfield
  {journal} {\bibinfo  {journal} {Phys. Rev. B}\ }\textbf {\bibinfo {volume}
  {85}},\ \bibinfo {pages} {045439} (\bibinfo {year} {2012})}\BibitemShut
  {NoStop}%
\bibitem [{\citenamefont {Christensen}\ and\ \citenamefont {von
  Lilienfeld}(2020)}]{christensen2020a}%
  \BibitemOpen
  \bibfield  {author} {\bibinfo {author} {\bibfnamefont {A.~S.}\ \bibnamefont
  {Christensen}}\ and\ \bibinfo {author} {\bibfnamefont {O.~A.}\ \bibnamefont
  {von Lilienfeld}},\ }\href {\doibase 10.1088/2632-2153/abba6f} {\bibfield
  {journal} {\bibinfo  {journal} {Mach. Learn.: Sci. Technol.}\ }\textbf
  {\bibinfo {volume} {1}},\ \bibinfo {pages} {045018} (\bibinfo {year}
  {2020})}\BibitemShut {NoStop}%
\bibitem [{\citenamefont {Heese}\ \emph {et~al.}(2017)\citenamefont {Heese},
  \citenamefont {Wolf}, \citenamefont {Dutrey},\ and\ \citenamefont
  {Guilloteau}}]{heese2017}%
  \BibitemOpen
  \bibfield  {author} {\bibinfo {author} {\bibfnamefont {S.}~\bibnamefont
  {Heese}}, \bibinfo {author} {\bibfnamefont {S.}~\bibnamefont {Wolf}},
  \bibinfo {author} {\bibfnamefont {A.}~\bibnamefont {Dutrey}}, \ and\ \bibinfo
  {author} {\bibfnamefont {S.}~\bibnamefont {Guilloteau}},\ }\href {\doibase
  10.1051/0004-6361/201730501} {\bibfield  {journal} {\bibinfo  {journal}
  {A\&A}\ }\textbf {\bibinfo {volume} {604}},\ \bibinfo {pages} {A5} (\bibinfo
  {year} {2017})}\BibitemShut {NoStop}%
\bibitem [{\citenamefont {Kolsbjerg}, \citenamefont {Peterson},\ and\
  \citenamefont {Hammer}(2018)}]{kolsbjerg2018}%
  \BibitemOpen
  \bibfield  {author} {\bibinfo {author} {\bibfnamefont {E.~L.}\ \bibnamefont
  {Kolsbjerg}}, \bibinfo {author} {\bibfnamefont {A.~A.}\ \bibnamefont
  {Peterson}}, \ and\ \bibinfo {author} {\bibfnamefont {B.}~\bibnamefont
  {Hammer}},\ }\href {\doibase 10.1103/PhysRevB.97.195424} {\bibfield
  {journal} {\bibinfo  {journal} {Phys. Rev. B}\ }\textbf {\bibinfo {volume}
  {97}},\ \bibinfo {pages} {195424} (\bibinfo {year} {2018})}\BibitemShut
  {NoStop}%
\bibitem [{\citenamefont {{van der Oord}}\ \emph {et~al.}(2022)\citenamefont
  {{van der Oord}}, \citenamefont {Sachs}, \citenamefont {Kov{\'a}cs},
  \citenamefont {Ortner},\ and\ \citenamefont {Cs{\'a}nyi}}]{vanderoord2022}%
  \BibitemOpen
  \bibfield  {author} {\bibinfo {author} {\bibfnamefont {C.}~\bibnamefont {{van
  der Oord}}}, \bibinfo {author} {\bibfnamefont {M.}~\bibnamefont {Sachs}},
  \bibinfo {author} {\bibfnamefont {D.~P.}\ \bibnamefont {Kov{\'a}cs}},
  \bibinfo {author} {\bibfnamefont {C.}~\bibnamefont {Ortner}}, \ and\ \bibinfo
  {author} {\bibfnamefont {G.}~\bibnamefont {Cs{\'a}nyi}},\ }\href {\doibase
  10.48550/arXiv.2210.04225} {\enquote {\bibinfo {title} {Hyperactive
  {{Learning}} ({{HAL}}) for {{Data-Driven Interatomic Potentials}}},}\ }
  (\bibinfo {year} {2022}),\ \Eprint {http://arxiv.org/abs/2210.04225}
  {arXiv:2210.04225 [physics, stat]} \BibitemShut {NoStop}%
\bibitem [{\citenamefont {Behler}(2015)}]{behler2015}%
  \BibitemOpen
  \bibfield  {author} {\bibinfo {author} {\bibfnamefont {J.}~\bibnamefont
  {Behler}},\ }\href {\doibase 10.1002/qua.24890} {\bibfield  {journal}
  {\bibinfo  {journal} {Int. J. Quantum Chem.}\ }\textbf {\bibinfo {volume}
  {115}},\ \bibinfo {pages} {1032} (\bibinfo {year} {2015})}\BibitemShut
  {NoStop}%
\bibitem [{\citenamefont {Zaverkin}, \citenamefont {Holzm{\"u}ller},\ and\
  \citenamefont {K{\"a}stner}(2021)}]{GM-NN-code}%
  \BibitemOpen
  \bibfield  {author} {\bibinfo {author} {\bibfnamefont {V.}~\bibnamefont
  {Zaverkin}}, \bibinfo {author} {\bibfnamefont {D.}~\bibnamefont
  {Holzm{\"u}ller}}, \ and\ \bibinfo {author} {\bibfnamefont {J.}~\bibnamefont
  {K{\"a}stner}},\ }\href@noop {} {\enquote {\bibinfo {title} {The {{Gaussian
  Moment Neural Network Package}}},}\ }\bibinfo {howpublished}
  {https://gitlab.com/zaverkin\_v/gmnn} (\bibinfo {year} {accessed 21 November
  2021})\BibitemShut {NoStop}%
\bibitem [{\citenamefont {Sch{\"u}tt}\ \emph {et~al.}(2019)\citenamefont
  {Sch{\"u}tt}, \citenamefont {Kessel}, \citenamefont {Gastegger},
  \citenamefont {Nicoli}, \citenamefont {Tkatchenko},\ and\ \citenamefont
  {M{\"u}ller}}]{SchNetPack}%
  \BibitemOpen
  \bibfield  {author} {\bibinfo {author} {\bibfnamefont {K.~T.}\ \bibnamefont
  {Sch{\"u}tt}}, \bibinfo {author} {\bibfnamefont {P.}~\bibnamefont {Kessel}},
  \bibinfo {author} {\bibfnamefont {M.}~\bibnamefont {Gastegger}}, \bibinfo
  {author} {\bibfnamefont {K.~A.}\ \bibnamefont {Nicoli}}, \bibinfo {author}
  {\bibfnamefont {A.}~\bibnamefont {Tkatchenko}}, \ and\ \bibinfo {author}
  {\bibfnamefont {K.-R.}\ \bibnamefont {M{\"u}ller}},\ }\href {\doibase
  10.1021/acs.jctc.8b00908} {\bibfield  {journal} {\bibinfo  {journal} {J.
  Chem. Theory Comput.}\ }\textbf {\bibinfo {volume} {15}},\ \bibinfo {pages}
  {448} (\bibinfo {year} {2019})}\BibitemShut {NoStop}%
\bibitem [{\citenamefont {Gastegger}\ and\ \citenamefont
  {Marquetand}(2020)}]{gastegger2020}%
  \BibitemOpen
  \bibfield  {author} {\bibinfo {author} {\bibfnamefont {M.}~\bibnamefont
  {Gastegger}}\ and\ \bibinfo {author} {\bibfnamefont {P.}~\bibnamefont
  {Marquetand}},\ }in\ \href {\doibase 10.1007/978-3-030-40245-7_12} {\emph
  {\bibinfo {booktitle} {Machine {{Learning Meets Quantum Physics}}}}},\
  \bibinfo {editor} {edited by\ \bibinfo {editor} {\bibfnamefont {K.~T.}\
  \bibnamefont {Sch{\"u}tt}}, \bibinfo {editor} {\bibfnamefont
  {S.}~\bibnamefont {Chmiela}}, \bibinfo {editor} {\bibfnamefont {O.~A.}\
  \bibnamefont {{von Lilienfeld}}}, \bibinfo {editor} {\bibfnamefont
  {A.}~\bibnamefont {Tkatchenko}}, \bibinfo {editor} {\bibfnamefont
  {K.}~\bibnamefont {Tsuda}}, \ and\ \bibinfo {editor} {\bibfnamefont {K.-R.}\
  \bibnamefont {M{\"u}ller}}}\ (\bibinfo  {publisher} {{Springer International
  Publishing}},\ \bibinfo {address} {{Cham}},\ \bibinfo {year} {2020})\ pp.\
  \bibinfo {pages} {233--252}\BibitemShut {NoStop}%
\bibitem [{\citenamefont {Smith}\ \emph {et~al.}(2018)\citenamefont {Smith},
  \citenamefont {Nebgen}, \citenamefont {Lubbers}, \citenamefont {Isayev},\
  and\ \citenamefont {Roitberg}}]{smith2018}%
  \BibitemOpen
  \bibfield  {author} {\bibinfo {author} {\bibfnamefont {J.~S.}\ \bibnamefont
  {Smith}}, \bibinfo {author} {\bibfnamefont {B.}~\bibnamefont {Nebgen}},
  \bibinfo {author} {\bibfnamefont {N.}~\bibnamefont {Lubbers}}, \bibinfo
  {author} {\bibfnamefont {O.}~\bibnamefont {Isayev}}, \ and\ \bibinfo {author}
  {\bibfnamefont {A.~E.}\ \bibnamefont {Roitberg}},\ }\href {\doibase
  10.1063/1.5023802} {\bibfield  {journal} {\bibinfo  {journal} {J. Chem.
  Phys.}\ }\textbf {\bibinfo {volume} {148}},\ \bibinfo {pages} {241733}
  (\bibinfo {year} {2018})}\BibitemShut {NoStop}%
\bibitem [{\citenamefont {Bernstein}, \citenamefont {Cs{\'a}nyi},\ and\
  \citenamefont {Deringer}(2019)}]{bernstein2019}%
  \BibitemOpen
  \bibfield  {author} {\bibinfo {author} {\bibfnamefont {N.}~\bibnamefont
  {Bernstein}}, \bibinfo {author} {\bibfnamefont {G.}~\bibnamefont
  {Cs{\'a}nyi}}, \ and\ \bibinfo {author} {\bibfnamefont {V.~L.}\ \bibnamefont
  {Deringer}},\ }\href {\doibase 10.1038/s41524-019-0236-6} {\bibfield
  {journal} {\bibinfo  {journal} {npj Comput Mater}\ }\textbf {\bibinfo
  {volume} {5}},\ \bibinfo {pages} {1} (\bibinfo {year} {2019})}\BibitemShut
  {NoStop}%
\bibitem [{\citenamefont {Schran}, \citenamefont {Brezina},\ and\ \citenamefont
  {Marsalek}(2020)}]{schran2020}%
  \BibitemOpen
  \bibfield  {author} {\bibinfo {author} {\bibfnamefont {C.}~\bibnamefont
  {Schran}}, \bibinfo {author} {\bibfnamefont {K.}~\bibnamefont {Brezina}}, \
  and\ \bibinfo {author} {\bibfnamefont {O.}~\bibnamefont {Marsalek}},\ }\href
  {\doibase 10.1063/5.0016004} {\bibfield  {journal} {\bibinfo  {journal} {J.
  Chem. Phys.}\ }\textbf {\bibinfo {volume} {153}},\ \bibinfo {pages} {104105}
  (\bibinfo {year} {2020})}\BibitemShut {NoStop}%
\bibitem [{\citenamefont {Vandermause}\ \emph {et~al.}(2020)\citenamefont
  {Vandermause}, \citenamefont {Torrisi}, \citenamefont {Batzner},
  \citenamefont {Xie}, \citenamefont {Sun}, \citenamefont {Kolpak},\ and\
  \citenamefont {Kozinsky}}]{vandermause2020}%
  \BibitemOpen
  \bibfield  {author} {\bibinfo {author} {\bibfnamefont {J.}~\bibnamefont
  {Vandermause}}, \bibinfo {author} {\bibfnamefont {S.~B.}\ \bibnamefont
  {Torrisi}}, \bibinfo {author} {\bibfnamefont {S.}~\bibnamefont {Batzner}},
  \bibinfo {author} {\bibfnamefont {Y.}~\bibnamefont {Xie}}, \bibinfo {author}
  {\bibfnamefont {L.}~\bibnamefont {Sun}}, \bibinfo {author} {\bibfnamefont
  {A.~M.}\ \bibnamefont {Kolpak}}, \ and\ \bibinfo {author} {\bibfnamefont
  {B.}~\bibnamefont {Kozinsky}},\ }\href {\doibase 10.1038/s41524-020-0283-z}
  {\bibfield  {journal} {\bibinfo  {journal} {Npj Comput. Mater.}\ }\textbf
  {\bibinfo {volume} {6}},\ \bibinfo {pages} {1} (\bibinfo {year}
  {2020})}\BibitemShut {NoStop}%
\bibitem [{\citenamefont {Miksch}\ \emph {et~al.}(2021)\citenamefont {Miksch},
  \citenamefont {Morawietz}, \citenamefont {K{\"a}stner}, \citenamefont
  {Urban},\ and\ \citenamefont {Artrith}}]{miksch2021a}%
  \BibitemOpen
  \bibfield  {author} {\bibinfo {author} {\bibfnamefont {A.~M.}\ \bibnamefont
  {Miksch}}, \bibinfo {author} {\bibfnamefont {T.}~\bibnamefont {Morawietz}},
  \bibinfo {author} {\bibfnamefont {J.}~\bibnamefont {K{\"a}stner}}, \bibinfo
  {author} {\bibfnamefont {A.}~\bibnamefont {Urban}}, \ and\ \bibinfo {author}
  {\bibfnamefont {N.}~\bibnamefont {Artrith}},\ }\href {\doibase
  10.1088/2632-2153/abfd96} {\bibfield  {journal} {\bibinfo  {journal} {Mach.
  Learn.: Sci. Technol.}\ }\textbf {\bibinfo {volume} {2}},\ \bibinfo {pages}
  {031001} (\bibinfo {year} {2021})}\BibitemShut {NoStop}%
\bibitem [{\citenamefont {Xie}\ \emph {et~al.}(2021)\citenamefont {Xie},
  \citenamefont {Vandermause}, \citenamefont {Sun}, \citenamefont
  {Cepellotti},\ and\ \citenamefont {Kozinsky}}]{xie2021}%
  \BibitemOpen
  \bibfield  {author} {\bibinfo {author} {\bibfnamefont {Y.}~\bibnamefont
  {Xie}}, \bibinfo {author} {\bibfnamefont {J.}~\bibnamefont {Vandermause}},
  \bibinfo {author} {\bibfnamefont {L.}~\bibnamefont {Sun}}, \bibinfo {author}
  {\bibfnamefont {A.}~\bibnamefont {Cepellotti}}, \ and\ \bibinfo {author}
  {\bibfnamefont {B.}~\bibnamefont {Kozinsky}},\ }\href {\doibase
  10.1038/s41524-021-00510-y} {\bibfield  {journal} {\bibinfo  {journal} {npj
  Comput Mater}\ }\textbf {\bibinfo {volume} {7}},\ \bibinfo {pages} {1}
  (\bibinfo {year} {2021})}\BibitemShut {NoStop}%
\bibitem [{\citenamefont {Young}\ \emph {et~al.}(2021)\citenamefont {Young},
  \citenamefont {{Johnston-Wood}}, \citenamefont {Deringer},\ and\
  \citenamefont {Duarte}}]{young2021}%
  \BibitemOpen
  \bibfield  {author} {\bibinfo {author} {\bibfnamefont {T.~A.}\ \bibnamefont
  {Young}}, \bibinfo {author} {\bibfnamefont {T.}~\bibnamefont
  {{Johnston-Wood}}}, \bibinfo {author} {\bibfnamefont {V.~L.}\ \bibnamefont
  {Deringer}}, \ and\ \bibinfo {author} {\bibfnamefont {F.}~\bibnamefont
  {Duarte}},\ }\href {\doibase 10.1039/D1SC01825F} {\bibfield  {journal}
  {\bibinfo  {journal} {Chem. Sci.}\ }\textbf {\bibinfo {volume} {12}},\
  \bibinfo {pages} {10944} (\bibinfo {year} {2021})}\BibitemShut {NoStop}%
\bibitem [{\citenamefont {Braams}\ and\ \citenamefont
  {Bowman}(2009)}]{braams2009}%
  \BibitemOpen
  \bibfield  {author} {\bibinfo {author} {\bibfnamefont {B.~J.}\ \bibnamefont
  {Braams}}\ and\ \bibinfo {author} {\bibfnamefont {J.~M.}\ \bibnamefont
  {Bowman}},\ }\href {\doibase 10.1080/01442350903234923} {\bibfield  {journal}
  {\bibinfo  {journal} {Int Rev Phys Chem}\ }\textbf {\bibinfo {volume} {28}},\
  \bibinfo {pages} {577} (\bibinfo {year} {2009})}\BibitemShut {NoStop}%
\bibitem [{\citenamefont {Qu}, \citenamefont {Yu},\ and\ \citenamefont
  {Bowman}(2018)}]{qu2018}%
  \BibitemOpen
  \bibfield  {author} {\bibinfo {author} {\bibfnamefont {C.}~\bibnamefont
  {Qu}}, \bibinfo {author} {\bibfnamefont {Q.}~\bibnamefont {Yu}}, \ and\
  \bibinfo {author} {\bibfnamefont {J.~M.}\ \bibnamefont {Bowman}},\ }\href
  {\doibase 10.1146/annurev-physchem-050317-021139} {\bibfield  {journal}
  {\bibinfo  {journal} {Annu. Rev. Phys. Chem.}\ }\textbf {\bibinfo {volume}
  {69}},\ \bibinfo {pages} {151} (\bibinfo {year} {2018})}\BibitemShut
  {NoStop}%
\bibitem [{\citenamefont {Bart{\'o}k}\ \emph {et~al.}(2017)\citenamefont
  {Bart{\'o}k}, \citenamefont {De}, \citenamefont {Poelking}, \citenamefont
  {Bernstein}, \citenamefont {Kermode}, \citenamefont {Cs{\'a}nyi},\ and\
  \citenamefont {Ceriotti}}]{bartok2017}%
  \BibitemOpen
  \bibfield  {author} {\bibinfo {author} {\bibfnamefont {A.~P.}\ \bibnamefont
  {Bart{\'o}k}}, \bibinfo {author} {\bibfnamefont {S.}~\bibnamefont {De}},
  \bibinfo {author} {\bibfnamefont {C.}~\bibnamefont {Poelking}}, \bibinfo
  {author} {\bibfnamefont {N.}~\bibnamefont {Bernstein}}, \bibinfo {author}
  {\bibfnamefont {J.~R.}\ \bibnamefont {Kermode}}, \bibinfo {author}
  {\bibfnamefont {G.}~\bibnamefont {Cs{\'a}nyi}}, \ and\ \bibinfo {author}
  {\bibfnamefont {M.}~\bibnamefont {Ceriotti}},\ }\href {\doibase
  10.1126/sciadv.1701816} {\bibfield  {journal} {\bibinfo  {journal} {Sci Adv}\
  }\textbf {\bibinfo {volume} {3}},\ \bibinfo {pages} {e1701816} (\bibinfo
  {year} {2017})}\BibitemShut {NoStop}%
\bibitem [{\citenamefont {Imbalzano}\ \emph {et~al.}(2018)\citenamefont
  {Imbalzano}, \citenamefont {Anelli}, \citenamefont {Giofr{\'e}},
  \citenamefont {Klees}, \citenamefont {Behler},\ and\ \citenamefont
  {Ceriotti}}]{imbalzano2018}%
  \BibitemOpen
  \bibfield  {author} {\bibinfo {author} {\bibfnamefont {G.}~\bibnamefont
  {Imbalzano}}, \bibinfo {author} {\bibfnamefont {A.}~\bibnamefont {Anelli}},
  \bibinfo {author} {\bibfnamefont {D.}~\bibnamefont {Giofr{\'e}}}, \bibinfo
  {author} {\bibfnamefont {S.}~\bibnamefont {Klees}}, \bibinfo {author}
  {\bibfnamefont {J.}~\bibnamefont {Behler}}, \ and\ \bibinfo {author}
  {\bibfnamefont {M.}~\bibnamefont {Ceriotti}},\ }\href {\doibase
  10.1063/1.5024611} {\bibfield  {journal} {\bibinfo  {journal} {J. Chem.
  Phys.}\ }\textbf {\bibinfo {volume} {148}},\ \bibinfo {pages} {241730}
  (\bibinfo {year} {2018})}\BibitemShut {NoStop}%
\bibitem [{\citenamefont {Schran}, \citenamefont {Behler},\ and\ \citenamefont
  {Marx}(2019)}]{schran2019}%
  \BibitemOpen
  \bibfield  {author} {\bibinfo {author} {\bibfnamefont {C.}~\bibnamefont
  {Schran}}, \bibinfo {author} {\bibfnamefont {J.}~\bibnamefont {Behler}}, \
  and\ \bibinfo {author} {\bibfnamefont {D.}~\bibnamefont {Marx}},\ }\href
  {\doibase 10.1021/acs.jctc.9b00805} {\bibfield  {journal} {\bibinfo
  {journal} {J. Chem. Theory Comput.}\ } (\bibinfo {year} {2019}),\
  10.1021/acs.jctc.9b00805}\BibitemShut {NoStop}%
\bibitem [{\citenamefont {Berendsen}\ \emph {et~al.}(1984)\citenamefont
  {Berendsen}, \citenamefont {Postma}, \citenamefont {{van Gunsteren}},
  \citenamefont {DiNola},\ and\ \citenamefont {Haak}}]{berendsen1984}%
  \BibitemOpen
  \bibfield  {author} {\bibinfo {author} {\bibfnamefont {H.~J.~C.}\
  \bibnamefont {Berendsen}}, \bibinfo {author} {\bibfnamefont {J.~P.~M.}\
  \bibnamefont {Postma}}, \bibinfo {author} {\bibfnamefont {W.~F.}\
  \bibnamefont {{van Gunsteren}}}, \bibinfo {author} {\bibfnamefont
  {A.}~\bibnamefont {DiNola}}, \ and\ \bibinfo {author} {\bibfnamefont {J.~R.}\
  \bibnamefont {Haak}},\ }\href {\doibase 10.1063/1.448118} {\bibfield
  {journal} {\bibinfo  {journal} {J. Chem. Phys.}\ }\textbf {\bibinfo {volume}
  {81}},\ \bibinfo {pages} {3684} (\bibinfo {year} {1984})}\BibitemShut
  {NoStop}%
\bibitem [{\citenamefont {Larsen}\ \emph {et~al.}(2017)\citenamefont {Larsen},
  \citenamefont {Mortensen}, \citenamefont {Blomqvist}, \citenamefont
  {Castelli}, \citenamefont {Christensen}, \citenamefont {{Marcin Du\l ak}},
  \citenamefont {Friis}, \citenamefont {Groves}, \citenamefont {Hammer},
  \citenamefont {Hargus}, \citenamefont {Hermes}, \citenamefont {Jennings},
  \citenamefont {Jensen}, \citenamefont {Kermode}, \citenamefont {Kitchin},
  \citenamefont {Kolsbjerg}, \citenamefont {Kubal}, \citenamefont {{Kristen
  Kaasbjerg}}, \citenamefont {Lysgaard}, \citenamefont {Maronsson},
  \citenamefont {Maxson}, \citenamefont {Olsen}, \citenamefont {Pastewka},
  \citenamefont {{Andrew Peterson}}, \citenamefont {Rostgaard}, \citenamefont
  {Schi{\o}tz}, \citenamefont {Sch{\"u}tt}, \citenamefont {Strange},
  \citenamefont {Thygesen}, \citenamefont {{Tejs Vegge}}, \citenamefont
  {Vilhelmsen}, \citenamefont {Walter}, \citenamefont {Zeng},\ and\
  \citenamefont {Jacobsen}}]{ASE}%
  \BibitemOpen
  \bibfield  {author} {\bibinfo {author} {\bibfnamefont {A.~H.}\ \bibnamefont
  {Larsen}}, \bibinfo {author} {\bibfnamefont {J.~J.}\ \bibnamefont
  {Mortensen}}, \bibinfo {author} {\bibfnamefont {J.}~\bibnamefont
  {Blomqvist}}, \bibinfo {author} {\bibfnamefont {I.~E.}\ \bibnamefont
  {Castelli}}, \bibinfo {author} {\bibfnamefont {R.}~\bibnamefont
  {Christensen}}, \bibinfo {author} {\bibnamefont {{Marcin Du\l ak}}}, \bibinfo
  {author} {\bibfnamefont {J.}~\bibnamefont {Friis}}, \bibinfo {author}
  {\bibfnamefont {M.~N.}\ \bibnamefont {Groves}}, \bibinfo {author}
  {\bibfnamefont {B.}~\bibnamefont {Hammer}}, \bibinfo {author} {\bibfnamefont
  {C.}~\bibnamefont {Hargus}}, \bibinfo {author} {\bibfnamefont {E.~D.}\
  \bibnamefont {Hermes}}, \bibinfo {author} {\bibfnamefont {P.~C.}\
  \bibnamefont {Jennings}}, \bibinfo {author} {\bibfnamefont {P.~B.}\
  \bibnamefont {Jensen}}, \bibinfo {author} {\bibfnamefont {J.}~\bibnamefont
  {Kermode}}, \bibinfo {author} {\bibfnamefont {J.~R.}\ \bibnamefont
  {Kitchin}}, \bibinfo {author} {\bibfnamefont {E.~L.}\ \bibnamefont
  {Kolsbjerg}}, \bibinfo {author} {\bibfnamefont {J.}~\bibnamefont {Kubal}},
  \bibinfo {author} {\bibnamefont {{Kristen Kaasbjerg}}}, \bibinfo {author}
  {\bibfnamefont {S.}~\bibnamefont {Lysgaard}}, \bibinfo {author}
  {\bibfnamefont {J.~B.}\ \bibnamefont {Maronsson}}, \bibinfo {author}
  {\bibfnamefont {T.}~\bibnamefont {Maxson}}, \bibinfo {author} {\bibfnamefont
  {T.}~\bibnamefont {Olsen}}, \bibinfo {author} {\bibfnamefont
  {L.}~\bibnamefont {Pastewka}}, \bibinfo {author} {\bibnamefont {{Andrew
  Peterson}}}, \bibinfo {author} {\bibfnamefont {C.}~\bibnamefont {Rostgaard}},
  \bibinfo {author} {\bibfnamefont {J.}~\bibnamefont {Schi{\o}tz}}, \bibinfo
  {author} {\bibfnamefont {O.}~\bibnamefont {Sch{\"u}tt}}, \bibinfo {author}
  {\bibfnamefont {M.}~\bibnamefont {Strange}}, \bibinfo {author} {\bibfnamefont
  {K.~S.}\ \bibnamefont {Thygesen}}, \bibinfo {author} {\bibnamefont {{Tejs
  Vegge}}}, \bibinfo {author} {\bibfnamefont {L.}~\bibnamefont {Vilhelmsen}},
  \bibinfo {author} {\bibfnamefont {M.}~\bibnamefont {Walter}}, \bibinfo
  {author} {\bibfnamefont {Z.}~\bibnamefont {Zeng}}, \ and\ \bibinfo {author}
  {\bibfnamefont {K.~W.}\ \bibnamefont {Jacobsen}},\ }\href {\doibase
  10.1088/1361-648X/aa680e} {\bibfield  {journal} {\bibinfo  {journal} {J.
  Phys.: Condens. Matter}\ }\textbf {\bibinfo {volume} {29}},\ \bibinfo {pages}
  {273002} (\bibinfo {year} {2017})}\BibitemShut {NoStop}%
\bibitem [{\citenamefont {Martyna}, \citenamefont {Klein},\ and\ \citenamefont
  {Tuckerman}(1992)}]{martyna1992}%
  \BibitemOpen
  \bibfield  {author} {\bibinfo {author} {\bibfnamefont {G.~J.}\ \bibnamefont
  {Martyna}}, \bibinfo {author} {\bibfnamefont {M.~L.}\ \bibnamefont {Klein}},
  \ and\ \bibinfo {author} {\bibfnamefont {M.}~\bibnamefont {Tuckerman}},\
  }\href {\doibase 10.1063/1.463940} {\bibfield  {journal} {\bibinfo  {journal}
  {J. Chem. Phys.}\ }\textbf {\bibinfo {volume} {97}},\ \bibinfo {pages} {2635}
  (\bibinfo {year} {1992})}\BibitemShut {NoStop}%
\bibitem [{\citenamefont {Thomas}\ \emph {et~al.}(2013)\citenamefont {Thomas},
  \citenamefont {Brehm}, \citenamefont {Fligg}, \citenamefont {V{\"o}hringer},\
  and\ \citenamefont {Kirchner}}]{thomas2013}%
  \BibitemOpen
  \bibfield  {author} {\bibinfo {author} {\bibfnamefont {M.}~\bibnamefont
  {Thomas}}, \bibinfo {author} {\bibfnamefont {M.}~\bibnamefont {Brehm}},
  \bibinfo {author} {\bibfnamefont {R.}~\bibnamefont {Fligg}}, \bibinfo
  {author} {\bibfnamefont {P.}~\bibnamefont {V{\"o}hringer}}, \ and\ \bibinfo
  {author} {\bibfnamefont {B.}~\bibnamefont {Kirchner}},\ }\href {\doibase
  10.1039/C3CP44302G} {\bibfield  {journal} {\bibinfo  {journal} {Phys. Chem.
  Chem. Phys.}\ }\textbf {\bibinfo {volume} {15}},\ \bibinfo {pages} {6608}
  (\bibinfo {year} {2013})}\BibitemShut {NoStop}%
\bibitem [{\citenamefont {Wiener}(1930)}]{wiener1930}%
  \BibitemOpen
  \bibfield  {author} {\bibinfo {author} {\bibfnamefont {N.}~\bibnamefont
  {Wiener}},\ }\href {\doibase 10.1007/BF02546511} {\bibfield  {journal}
  {\bibinfo  {journal} {Acta Math.}\ }\textbf {\bibinfo {volume} {55}},\
  \bibinfo {pages} {117} (\bibinfo {year} {1930})}\BibitemShut {NoStop}%
\bibitem [{\citenamefont {Blackman}\ and\ \citenamefont
  {Tukey}(1958)}]{blackman1958}%
  \BibitemOpen
  \bibfield  {author} {\bibinfo {author} {\bibfnamefont {R.~B.}\ \bibnamefont
  {Blackman}}\ and\ \bibinfo {author} {\bibfnamefont {J.~W.}\ \bibnamefont
  {Tukey}},\ }\href {\doibase 10.1002/j.1538-7305.1958.tb03874.x} {\bibfield
  {journal} {\bibinfo  {journal} {Bell Labs Tech. J.}\ }\textbf {\bibinfo
  {volume} {37}},\ \bibinfo {pages} {185} (\bibinfo {year} {1958})}\BibitemShut
  {NoStop}%
\bibitem [{\citenamefont {Neese}\ \emph {et~al.}(2020)\citenamefont {Neese},
  \citenamefont {Wennmohs}, \citenamefont {Becker},\ and\ \citenamefont
  {Riplinger}}]{ORCA5}%
  \BibitemOpen
  \bibfield  {author} {\bibinfo {author} {\bibfnamefont {F.}~\bibnamefont
  {Neese}}, \bibinfo {author} {\bibfnamefont {F.}~\bibnamefont {Wennmohs}},
  \bibinfo {author} {\bibfnamefont {U.}~\bibnamefont {Becker}}, \ and\ \bibinfo
  {author} {\bibfnamefont {C.}~\bibnamefont {Riplinger}},\ }\href {\doibase
  10.1063/5.0004608} {\bibfield  {journal} {\bibinfo  {journal} {J. Chem.
  Phys.}\ }\textbf {\bibinfo {volume} {152}},\ \bibinfo {pages} {224108}
  (\bibinfo {year} {2020})}\BibitemShut {NoStop}%
\bibitem [{\citenamefont {Adamo}\ and\ \citenamefont {Barone}(1999)}]{PBE0}%
  \BibitemOpen
  \bibfield  {author} {\bibinfo {author} {\bibfnamefont {C.}~\bibnamefont
  {Adamo}}\ and\ \bibinfo {author} {\bibfnamefont {V.}~\bibnamefont {Barone}},\
  }\href {\doibase 10.1063/1.478522} {\bibfield  {journal} {\bibinfo  {journal}
  {J. Chem. Phys.}\ }\textbf {\bibinfo {volume} {110}},\ \bibinfo {pages}
  {6158} (\bibinfo {year} {1999})}\BibitemShut {NoStop}%
\bibitem [{\citenamefont {Weigend}\ and\ \citenamefont
  {Ahlrichs}(2005)}]{def2}%
  \BibitemOpen
  \bibfield  {author} {\bibinfo {author} {\bibfnamefont {F.}~\bibnamefont
  {Weigend}}\ and\ \bibinfo {author} {\bibfnamefont {R.}~\bibnamefont
  {Ahlrichs}},\ }\href {\doibase 10.1039/B508541A} {\bibfield  {journal}
  {\bibinfo  {journal} {Phys. Chem. Chem. Phys.}\ }\textbf {\bibinfo {volume}
  {7}},\ \bibinfo {pages} {3297} (\bibinfo {year} {2005})}\BibitemShut
  {NoStop}%
\bibitem [{\citenamefont {Weigend}(2006)}]{def2-fit}%
  \BibitemOpen
  \bibfield  {author} {\bibinfo {author} {\bibfnamefont {F.}~\bibnamefont
  {Weigend}},\ }\href {\doibase 10.1039/B515623H} {\bibfield  {journal}
  {\bibinfo  {journal} {Phys. Chem. Chem. Phys.}\ }\textbf {\bibinfo {volume}
  {8}},\ \bibinfo {pages} {1057} (\bibinfo {year} {2006})}\BibitemShut
  {NoStop}%
\bibitem [{\citenamefont {Neese}\ \emph {et~al.}(2009)\citenamefont {Neese},
  \citenamefont {Wennmohs}, \citenamefont {Hansen},\ and\ \citenamefont
  {Becker}}]{RIJCOSX}%
  \BibitemOpen
  \bibfield  {author} {\bibinfo {author} {\bibfnamefont {F.}~\bibnamefont
  {Neese}}, \bibinfo {author} {\bibfnamefont {F.}~\bibnamefont {Wennmohs}},
  \bibinfo {author} {\bibfnamefont {A.}~\bibnamefont {Hansen}}, \ and\ \bibinfo
  {author} {\bibfnamefont {U.}~\bibnamefont {Becker}},\ }\href {\doibase
  10.1016/j.chemphys.2008.10.036} {\bibfield  {journal} {\bibinfo  {journal}
  {Chem. Phys.}\ }\textbf {\bibinfo {volume} {356}},\ \bibinfo {pages} {98}
  (\bibinfo {year} {2009})}\BibitemShut {NoStop}%
\bibitem [{\citenamefont {Neese}(2003)}]{neese2003}%
  \BibitemOpen
  \bibfield  {author} {\bibinfo {author} {\bibfnamefont {F.}~\bibnamefont
  {Neese}},\ }\href {\doibase 10.1002/jcc.10318} {\bibfield  {journal}
  {\bibinfo  {journal} {J. Comput. Chem.}\ }\textbf {\bibinfo {volume} {24}},\
  \bibinfo {pages} {1740} (\bibinfo {year} {2003})}\BibitemShut {NoStop}%
\bibitem [{\citenamefont {Zeegers}\ \emph {et~al.}(2023)\citenamefont
  {Zeegers}, \citenamefont {Guiu}, \citenamefont {Kemper}, \citenamefont
  {Marshall},\ and\ \citenamefont {Bromley}}]{zeegers2023}%
  \BibitemOpen
  \bibfield  {author} {\bibinfo {author} {\bibfnamefont {S.}~\bibnamefont
  {Zeegers}}, \bibinfo {author} {\bibfnamefont {J.~M.}\ \bibnamefont {Guiu}},
  \bibinfo {author} {\bibfnamefont {F.}~\bibnamefont {Kemper}}, \bibinfo
  {author} {\bibfnamefont {J.}~\bibnamefont {Marshall}}, \ and\ \bibinfo
  {author} {\bibfnamefont {S.~T.}\ \bibnamefont {Bromley}},\ }\href {\doibase
  10.1039/D3FD00055A} {\bibfield  {journal} {\bibinfo  {journal} {Faraday
  Discuss.}\ } (\bibinfo {year} {2023}),\ 10.1039/D3FD00055A}\BibitemShut
  {NoStop}%
\bibitem [{\citenamefont {Bart{\'o}k}, \citenamefont {Kondor},\ and\
  \citenamefont {Cs{\'a}nyi}(2013)}]{SOAP}%
  \BibitemOpen
  \bibfield  {author} {\bibinfo {author} {\bibfnamefont {A.~P.}\ \bibnamefont
  {Bart{\'o}k}}, \bibinfo {author} {\bibfnamefont {R.}~\bibnamefont {Kondor}},
  \ and\ \bibinfo {author} {\bibfnamefont {G.}~\bibnamefont {Cs{\'a}nyi}},\
  }\href {\doibase 10.1103/PhysRevB.87.184115} {\bibfield  {journal} {\bibinfo
  {journal} {Phys. Rev. B}\ }\textbf {\bibinfo {volume} {87}},\ \bibinfo
  {pages} {184115} (\bibinfo {year} {2013})}\BibitemShut {NoStop}%
\bibitem [{\citenamefont {Zeegers}\ \emph {et~al.}(2021)\citenamefont
  {Zeegers}, \citenamefont {Bouwman}, \citenamefont {Chiar}, \citenamefont
  {Costantini}, \citenamefont {Decleir}, \citenamefont {Dharmawardena},
  \citenamefont {Geballe}, \citenamefont {Gordon}, \citenamefont {Green},
  \citenamefont {Henning}, \citenamefont {Jones}, \citenamefont {Kemper},
  \citenamefont {Li}, \citenamefont {Marshall}, \citenamefont {McClure},
  \citenamefont {Misselt}, \citenamefont {Pendleton}, \citenamefont {Perotti},
  \citenamefont {Pontoppidan}, \citenamefont {Potapov}, \citenamefont
  {Scicluna}, \citenamefont {Tielens}, \citenamefont {Waters},\ and\
  \citenamefont {Zari}}]{zeegers2021}%
  \BibitemOpen
  \bibfield  {author} {\bibinfo {author} {\bibfnamefont {S.}~\bibnamefont
  {Zeegers}}, \bibinfo {author} {\bibfnamefont {J.}~\bibnamefont {Bouwman}},
  \bibinfo {author} {\bibfnamefont {J.}~\bibnamefont {Chiar}}, \bibinfo
  {author} {\bibfnamefont {E.}~\bibnamefont {Costantini}}, \bibinfo {author}
  {\bibfnamefont {M.}~\bibnamefont {Decleir}}, \bibinfo {author} {\bibfnamefont
  {T.}~\bibnamefont {Dharmawardena}}, \bibinfo {author} {\bibfnamefont {T.~R.}\
  \bibnamefont {Geballe}}, \bibinfo {author} {\bibfnamefont {K.~D.}\
  \bibnamefont {Gordon}}, \bibinfo {author} {\bibfnamefont {J.~D.}\
  \bibnamefont {Green}}, \bibinfo {author} {\bibfnamefont {T.~K.}\ \bibnamefont
  {Henning}}, \bibinfo {author} {\bibfnamefont {O.}~\bibnamefont {Jones}},
  \bibinfo {author} {\bibfnamefont {F.}~\bibnamefont {Kemper}}, \bibinfo
  {author} {\bibfnamefont {A.}~\bibnamefont {Li}}, \bibinfo {author}
  {\bibfnamefont {J.~P.}\ \bibnamefont {Marshall}}, \bibinfo {author}
  {\bibfnamefont {M.}~\bibnamefont {McClure}}, \bibinfo {author} {\bibfnamefont
  {K.}~\bibnamefont {Misselt}}, \bibinfo {author} {\bibfnamefont {Y.~J.}\
  \bibnamefont {Pendleton}}, \bibinfo {author} {\bibfnamefont {G.}~\bibnamefont
  {Perotti}}, \bibinfo {author} {\bibfnamefont {K.~M.}\ \bibnamefont
  {Pontoppidan}}, \bibinfo {author} {\bibfnamefont {A.}~\bibnamefont
  {Potapov}}, \bibinfo {author} {\bibfnamefont {P.}~\bibnamefont {Scicluna}},
  \bibinfo {author} {\bibfnamefont {A.}~\bibnamefont {Tielens}}, \bibinfo
  {author} {\bibfnamefont {R.}~\bibnamefont {Waters}}, \ and\ \bibinfo {author}
  {\bibfnamefont {E.}~\bibnamefont {Zari}},\ }\href
  {https://ui.adsabs.harvard.edu/abs/2021jwst.prop.2183Z} {\bibfield  {journal}
  {\bibinfo  {journal} {JWST Proposal. Cycle 1}\ ,\ \bibinfo {pages} {2183}}
  (\bibinfo {year} {2021})}\BibitemShut {NoStop}%
\end{thebibliography}%

\end{document}